\documentclass[aps,prd,showpacs,twocolumn,10pt,superscriptaddress,floatfix,nofootinbib,longbibliography]{revtex4-1}
\usepackage{amssymb,amsfonts,amsmath,bm,bbm}
\usepackage{graphicx}
\usepackage{dcolumn}
\usepackage[
colorlinks=true,
linkcolor=blue,  
breaklinks=true,
urlcolor=blue,
citecolor=blue]{hyperref}
\usepackage{bbold}
\usepackage{epstopdf}
\usepackage{CJK}

\newcommand{\HNUST}{\affiliation{
Hunan Provincial Key Laboratory of Intelligent Sensors and Advanced Sensor Materials, \\ School of Physics and Electronics, Hunan University of Science and Technology, Xiangtan 411201, China}}

\newcommand{\HBUAT}{\affiliation{
Hubei Key Laboratory of Energy Storage and Power Battery, School of Optoelectronic Engineering, School of New Energy, Hubei University of Automotive Technology, Shiyan 442002, China}}
            
\newcommand{\HNUSTchem}{\affiliation{
School of Chemistry and Chemical Engineering, Hunan University of Science and Technology, Xiangtan 411201, China}}

\newcommand{\UCAS}{\affiliation{
School of Nuclear Science and Technology,
      University of Chinese Academy of Sciences,
      Beijing 100049, China}}
      
\newcommand{\IHEP}{\affiliation{
Institute of High Energy Physics, Chinese Academy of Sciences, Beijing 100049, China}}

\newcommand{\USC}{\affiliation{
School of Nuclear Science and Technology, University of South China, Hengyang 421001, China}}

\newcommand{\INFNch}{\affiliation{
INFN—Istituto Nazionale di Fisica Nucleare—Sezione di Bari, Via Orabona 4, 70125 Bari, Italy}}
      
\newcommand{\HNU}{\affiliation{
Synergetic Innovation Center for Quantum Effects and Application, Hunan Normal University, Changsha 410081, China}}

\begin{document}

\title{Lattice-data-driven specific heat and isentropic bulk modulus of SU(3) gluon matter at finite temperature}

\author{Wei Shen} 
\HNUST

\author{Zhen-Yan Lu}
\email[Corresponding author:~]
{luzhenyan@hnust.edu.cn}
\HNUST

\author{Muhammad Waqas} 
\HBUAT

\author{Xun Chen}\USC\INFNch

\author{Zhi-Jun Ma} \email{mazhijun@hnust.edu.cn}\HNUST\HNUSTchem

\author{Guang-Xiong Peng} \email{gxpeng@ucas.ac.cn}\UCAS\IHEP\HNU

\date{\today}

\begin{abstract}

We investigate the specific heat and isentropic bulk modulus of finite-temperature pure SU(3) gauge matter within a lattice-data-driven phenomenological framework. The equation of state is formulated in terms of a temperature-dependent effective gluon mass constrained { by lattice QCD pressure data as input, allowing the pressure}, trace anomaly, gluon number density, energy per thermally active gluonic mode, and derivative-sensitive response functions to be derived in a thermodynamically consistent manner.  
The resulting pressure and trace anomaly reproduce the characteristic lattice behavior across the deconfinement region, while the effective gluonic degrees of freedom increase rapidly above $T_c$. The normalized specific heat $C_V/T^3$ develops a pronounced enhancement in the vicinity of $T_c$,
reflecting the rapid temperature variation of the energy density across the deconfinement region. The isentropic bulk modulus $K_S/T^4$ also rises sharply across the transition region, indicating a substantial stiffening of the equation of state. At high temperatures, both response functions gradually approach values close to their massless conformal Stefan--Boltzmann reference values, 
with $\left(C_V/T^3\right)_{\rm SB}=32\pi^2/15\simeq 21.06$ and 
$\left(K_S/T^4\right)_{\rm SB}=32\pi^2/135\simeq 2.34$. These findings indicate that the specific heat and isentropic bulk modulus provide complementary constraints on the temperature evolution of nonconformal dynamics in pure SU(3) gauge matter.

\end{abstract}

\maketitle

\section{Introduction} \label{sec:INTRODUCTION}

The thermodynamic properties of strongly interacting matter at finite temperatures constitute a central topic in quantum chromodynamics (QCD)~\cite{Gross:2022hyw,Brambilla:2014jmp,MUSES:2023hyz,Adhikari:2024bfa,Bresciani:2025mcu,Chen:2024aom},  
with direct relevance to the formation and evolution of the quark–gluon plasma in relativistic heavy-ion collisions~\cite{Gyulassy:2004zy,PHENIX:2004vcz,STAR:2005gfr,BRAHMS:2004adc,Shuryak:2004cy,Harris:2023tti,Heinz:2013th,Shou:2024uga,Aarts:2023vsf,ALICE:2022hor,Ratti:2018ksb,Chen:2024epd,Altmann:2024icx}. 
In particular, the thermodynamic properties of the quark–gluon plasma, including its equation of state and response functions, play a key role in controlling its collective behavior, hydrodynamic evolution, transport coefficients, and fluctuations~\cite{He:2024amc,Ding:2026gco,Astrakhantsev:2019zkr,Borsanyi:2018grb,Puglisi:2014pda,Zhao:2017vmx,Parisi:2025gwq}.  
First-principles lattice QCD simulations have established the equation of state of the quark–gluon plasma at finite temperature and vanishing chemical potential with high precision~\cite{Bazavov:2017dsy,Borsanyi:2013bia,Aoki:2005vt,Borsanyi:2012cr,Borsanyi:2010cj,Caselle:2018kap}, as well as that of pure SU(3) gauge theory over a wide temperature range~\cite{Borsanyi:2012ve,Giusti:2016iqr,Asakawa:2013laa}. These studies reveal pronounced deviations from conformal behavior in the vicinity of the deconfinement temperature $T_c$, followed by a gradual approach toward approximate scale invariance at higher temperatures~\cite{Boyd:1996bx,Giusti:2025fxu}.
In parallel, quasiparticle, data-driven, and holographically inspired approaches have been developed to complement lattice results and provide analytical insights into QCD thermodynamics~\cite{Plumari:2011mk,Sasaki:2012bi,Ruggieri:2012ny,Castorina:2011ja,Heshmatian:2023yzz,Chen:2025goz,He:2022amv,Chen:2025kqb,Yaresko:2013tia,Alba:2014lda}.

Bulk thermodynamic quantities, such as the pressure, energy density, entropy density, and trace anomaly, characterize the global features of the equation of state~\cite{Ding:2015ona,HotQCD:2014kol}. However, these quantities do not fully encode the response of the medium to external perturbations. A more refined understanding of hot QCD matter requires thermodynamic response
functions and susceptibilities, which quantify how the system reacts to variations
in temperature, pressure, and density~\cite{Lu:2025cls,Fu:2016tey,Braun-Munzinger:2020jbk,Sorensen:2021zme,Lu:2019diy}. In particular, observables such as the specific heat and isentropic bulk modulus probe the stiffness of the equation of state, the degree of scale breaking, and the nature of thermodynamic fluctuations near the deconfinement transition~\cite{Karsch:2015nqx,Gavai:2004se,Borsanyi:2022soo,HotQCD:2018pds,Borsanyi:2020fev}. These quantities are also directly relevant for the hydrodynamic description of the quark–gluon plasma, influencing the propagation of sound modes and density fluctuations~\cite{Romatschke:2007mq,Alba:2017mqu}.

Despite the high precision achieved in lattice determinations of the equation of state, a systematic understanding of higher-order thermodynamic response functions in pure gluonic matter remains incomplete. 
While lattice calculations provide accurate bulk thermodynamic quantities, 
the extraction of higher-order derivatives is typically subject to numerical uncertainties and does not directly reveal the underlying microscopic dynamics. In particular, the temperature dependence of second-order response functions across the deconfinement region, and their connection to nonconformal dynamics, remain insufficiently understood within a unified framework constrained by lattice data. In previous work, we developed a lattice-data-driven phenomenological framework for pure SU(3) gluon matter based on a temperature-dependent effective gluon mass~\cite{Ma:2026xxx}. 
In this work, we refer to this approach as the temperature-dependent mass (TDM) model, in which the temperature dependence of the effective gluon mass is incorporated to encode medium effects. 
In this approach, the effective mass is constrained by {lattice pressure data} and encodes medium effects in a transparent manner~\cite{Chen:2021edy,Chen:2023rza}. This framework reproduces bulk thermodynamic observables over a wide temperature range and captures essential features of nonconformal dynamics.

Motivated by these observations, it is natural to investigate whether higher-order response functions exhibit similarly robust and physically interpretable structures. 
The present work extends this lattice-constrained framework~\cite{Ma:2026xxx} within the TDM model to a systematic study of second-order thermodynamic response functions, 
focusing on the specific heat and the isentropic bulk modulus of pure SU(3) gluon matter. 
As second derivatives of the thermodynamic potential, these observables are directly related to fluctuations and susceptibilities, 
and thus provide a more sensitive probe of the underlying medium properties. 
The specific heat characterizes the response of the energy density to temperature variations, 
while the isentropic bulk modulus quantifies the mechanical stiffness of the system through $(E+P)c_s^2$.
Our results show that near $T_c$ the rapid increase of gluon density and energy density is accompanied by enhanced energy fluctuations, leading to a pronounced enhancement in the specific heat, while the isentropic bulk modulus increases sharply, indicating a significant stiffening of the equation of state.
At higher temperatures, both observables gradually move toward their conformal reference behavior, consistent with the restoration of approximate scale invariance. These features reflect the interplay between deconfinement dynamics, the rapid variation of the effective gluon mass, and the associated breaking of scale invariance.

The paper is organized as follows. 
In Sec.~\ref{sec:TheoFrame}, we introduce the theoretical framework for pure SU(3) gluon matter with a temperature-dependent effective gluon mass. 
In particular, we formulate a thermodynamically consistent treatment of the system, derive the relevant thermodynamic relations, and specify the lattice-data-driven parametrization of the effective mass. 
In Sec.~\ref{sec:results}, we present our numerical results. 
We first discuss bulk thermodynamic quantities to establish the baseline properties of the system, and then analyze the second-order thermodynamic response functions, focusing on the specific heat and the isentropic bulk modulus, and their temperature dependence across the deconfinement transition. 
Finally, Sec.~\ref{sec:CONCLUSION} summarizes our main findings and provides an outlook.

\section{Thermodynamic framework with a temperature-dependent effective gluon mass} 
\label{sec:TheoFrame}

We study pure SU(3) gluon matter at finite temperature and vanishing chemical potential within the TDM model~\cite{Wen:2005uf,Xia:2014zaa,Lu:2016jsv,Issifu:2023qoo}. 
The central assumption of this approach is that medium effects and nonconformal interactions can be encoded, at the level of thermodynamics, through a temperature-dependent effective gluon mass $m_g(T)$. This quantity should be understood as an effective parameter entering the thermal dispersion relation, rather than as a gauge-invariant pole mass of a propagating gluon. Similar effective quasiparticle descriptions have been widely used to encode interaction effects in hot Yang--Mills/QCD thermodynamics~\cite{Plumari:2011mk,Ruggieri:2012ny,Sasaki:2012bi}, although the detailed temperature dependence of the effective mass or fugacity is generally model dependent~\cite{Gorenstein:1995vm,Peshier:1995ty,Levai:1997yx,Bannur:2005wm,Luo:2013bcz,Sambataro:2024mkr}. In particular, in the low-temperature regime it does not imply the existence of freely propagating massive gluons in the confined phase; instead, it parametrizes the suppression of colored gluonic degrees of freedom and provides an effective description of the thermodynamics inferred from lattice data. 
 
The motivation for introducing $m_g(T)$ is not to mimic a specific microscopic mass-generation mechanism, but to construct a minimal thermodynamic framework that reproduces the lattice equation of state and can be systematically extended to higher-order response functions. Since the dispersion relation depends implicitly on temperature through $m_g(T)$, thermodynamic consistency requires that both the explicit and implicit temperature dependences be treated on equal footing. Similar consistency requirements also arise in effective-mass descriptions of
strongly interacting matter, where medium-dependent masses or couplings modify
the thermodynamic derivatives~\cite{Ma:2023stj,Wang:2025lwv}. The importance of thermodynamic consistency in quasiparticle descriptions has also been emphasized in related effective approaches~\cite{Luo:2013bcz,Bannur:2007tk}. This issue is especially important when evaluating response functions involving temperature derivatives, such as the specific heat and the isentropic bulk modulus, which are more sensitive to the structure of the model than bulk quantities alone.

\subsection{Thermodynamically consistent formulation}

For a pure gluonic system, the thermodynamic state is specified by the temperature $T$ and the volume $V$. 
We begin with the grand thermodynamic potential $\bar{\Omega}$, whose differential form reads
\begin{equation}
d\bar{\Omega} = -\bar{S}\, dT - P\, dV - \bar{N}\, d\mu,
\end{equation}
where $\bar{S}$, $P$, $\mu$, and $\bar{N}$ denote the entropy, pressure, chemical potential, and particle number, respectively. For gluons, there is no conserved charge and thus $\mu = 0$, so that the Helmholtz free energy $\bar{F}$ coincides with the thermodynamic potential, $\bar{F} = \bar{\Omega}$. Introducing the corresponding densities,
$F = \bar{F}/V, \Omega_g = \bar{\Omega}/V, S = \bar{S}/V$,
one has
\begin{equation}
F(T,V,m_g) = \Omega_g(T,V,m_g).
\end{equation}
Taking the total differential, and noting that the effective gluon mass depends only on temperature, $m_g = m_g(T)$, we obtain
\begin{equation}
dF = d\Omega_g 
= \frac{\partial \Omega_g}{\partial T} dT 
+ \frac{\partial \Omega_g}{\partial m_g} dm_g
+ \frac{\partial \Omega_g}{\partial V} dV.
\end{equation}

To incorporate medium effects, we introduce a temperature-dependent effective gluon mass $m_g(T)$ as discussed in previous section. 
Since $m_g$ depends only on temperature,
\begin{equation}
dm_g = \frac{d m_g}{dT} dT,
\end{equation}
which leads to
\begin{equation}
dF = \left( \frac{\partial \Omega_g}{\partial T}
+ \frac{\partial \Omega_g}{\partial m_g} \frac{d m_g}{dT} \right) dT
+ \frac{\partial \Omega_g}{\partial V} dV.
\end{equation}

On the other hand, the thermodynamic identity for the free energy density reads
\begin{equation}
dF=-S\,dT-\left(P+F\right)\frac{dV}{V}.
\label{eq:dFdensity}
\end{equation}

By comparison, we identify the entropy density as
\begin{equation}
S = - \frac{\partial \Omega_g}{\partial T}
- \frac{\partial \Omega_g}{\partial m_g} \frac{d m_g}{dT},
\label{eq:entropy}
\end{equation}
and the pressure as
$P=-F-V\frac{\partial \Omega_g}{\partial V}$. 
For a homogeneous system in the thermodynamic limit, the density $\Omega_g$ is volume independent, i.e.,  $\partial \Omega_g/\partial V=0$, and the pressure then reduces to the familiar relation
\begin{equation}
P=-F=-\Omega_g.
\label{eq:Pbasic}
\end{equation}

The energy density follows from $E = F + TS$, 
which yields
\begin{equation}
E = \Omega_g 
- T \frac{\partial \Omega_g}{\partial T}
- T \frac{\partial \Omega_g}{\partial m_g} \frac{d m_g}{dT}.
\label{eq:energy}
\end{equation}
Equations~(\ref{eq:entropy})-(\ref{eq:energy}) define the thermodynamically consistent formulation of the TDM model. In the limit of a temperature-independent mass, they reduce to the standard expressions for an ideal bosonic gas.

\subsection{Thermodynamic quantities in the TDM model}

The thermodynamic potential density of the gluonic system is given by
\begin{equation}
\Omega_g
= d_g T\int \frac{d^3p}{(2\pi)^3} 
\ln \left(1 - e^{-E_p/T} \right),
\end{equation}
where $E_p=\sqrt{p^2+m_g^2(T)}$ and $d_g=16$ is the gluon degeneracy factor. 
Using Eq.~(\ref{eq:Pbasic}), the pressure takes the form
\begin{equation}
\frac{P_g}{T^4}
=
-\frac{d_g}{2\pi^2}
\int_0^{\infty}
\ln\left(1-e^{-\sqrt{y^2+(m_g/T)^2}}\right) y^2~\mathrm{d}y ,
\label{eq:Pgmg}
\end{equation}
where $y \equiv p/T$ is a dimensionless momentum variable. 
For $m_g/T\ll 1$, Eq.~(\ref{eq:Pgmg}) approaches the Stefan--Boltzmann result. In the present fitted parametrization, however, $m_g/T$ tends to a small nonzero constant at asymptotically high temperature, so the Stefan--Boltzmann value should be regarded as a conformal reference rather than an exact asymptote of the model.

For later discussion, it is useful to display the small-$m_g/T$ expansion of Eq.~(\ref{eq:Pgmg}). Defining $z\equiv m_g/(\pi T)$, one obtains
 \begin{eqnarray} \label{eq:PP0expgen}
\frac{P_g}{P_0}
&=&
 1-\frac{15}{4}z^2 +\frac{15}{2}z^3 +{\cal O}(z^4) ,
\end{eqnarray}
where $P_0\equiv (d_g/90)\pi^2T^4$. This expansion makes explicit how deviations from the massless conformal reference behavior are controlled by the ratio $m_g/T$.

The gluon number density, although not associated with a conserved charge, is still useful for characterizing the thermal population of modes and is given by
\begin{equation}
\frac{n_g}{T^3}
=
\frac{d_g}{2\pi^2}
\int_0^{\infty}
\frac{y^2\,\mathrm{d}y}
{e^{\sqrt{y^2+(m_g/T)^2}}-1}.
\label{eq:ng}
\end{equation}
Building on the temperature dependence of $P_g$ and $E_g$, higher-order thermodynamic response functions can be evaluated in a systematic manner. In this work, we focus on the specific heat and the isentropic bulk modulus, which involve second derivatives of the thermodynamic potential and are therefore particularly sensitive to variations of the effective gluon mass. These observables provide complementary insights into the stiffness of the equation of state and the degree of nonconformality, and their behavior across the deconfinement transition will be analyzed in detail in the following sections.

\subsection{Lattice-data-driven parametrization of the effective gluon mass}

{
To close the system, an explicit form of the effective gluon mass $m_g(T)$ is required. In the present TDM framework, $m_g(T)$ should be understood as an effective thermodynamic parameter rather than a microscopic pole mass of a propagating gluon. Its purpose is not to provide a first-principles derivation of mass generation, but rather to encode in a compact and thermodynamically consistent way the medium effects necessary to reproduce the lattice equation of state over a broad temperature range.

We now make the determination procedure of $m_g(T)$ explicit. In this work, the only lattice input is the normalized pressure $P/T^4$ of pure SU(3) gauge theory from Ref.~\cite{Borsanyi:2012ve}. No use is made of the energy density, trace anomaly, entropy density, or any combined thermodynamic fit. Instead, the effective mass profile is fully constrained by the pressure data alone. 
Specifically, for each reduced temperature 
$x_i \equiv T_i/T_c$, 
the corresponding value of the dimensionless effective mass, $q_i \equiv m_g(T_i)/T_i$, 
is obtained by numerically solving
\begin{eqnarray}
-\frac{d_g}{2\pi^2}
\int_0^\infty y^2
\ln\left(
1-e^{-\sqrt{y^2+q_i^2}}
\right)dy
=
\left(\frac{P}{T^4}\right)_{\rm lat}(x_i), ~~
\label{eq:pressure_matching}
\end{eqnarray}
where the left-hand side is the model expression derived from Eq.~(\ref{eq:Pgmg}). This procedure gives a set of 48 pressure-extracted values of $m_g/T$. These discrete values are then fitted separately below and above the deconfinement temperature.

For $x\leq 1$, $m_g/T$ is fitted as a cubic polynomial in $x$, while for $x>1$ it is fitted as a cubic polynomial in the running coupling $\alpha(x)$. Therefore, the entire parametrization is constrained solely by lattice pressure data, without introducing additional thermodynamic observables as independent fitting inputs. Once $m_g(T)$ is fixed, the energy density, entropy density, and trace anomaly are obtained from thermodynamically consistent relations in Eqs.~(\ref{eq:entropy})-(\ref{eq:energy}).  
Introducing the reduced temperature $x \equiv T/T_c$, the effective mass below the deconfinement temperature is parametrized as
\begin{equation}
\frac{m_g}{T}
=
a_0 + a_1 x + a_2 x^2 + a_3 x^3 ,
\qquad (T<T_c),
\label{eq:mg_lowT}
\end{equation}
with coefficients
$a_0 = 67.018$,
$a_1 = -189.089$,
$a_2 = 212.666$,
and
$a_3 = -83.605$.
This parametrization captures the strong suppression of effective colored gluonic excitations in the vicinity of the confined phase and the rapid change of thermodynamic behavior near $T_c$.

For temperatures above the critical point, the pressure-extracted mass profile is parametrized as a polynomial in the running coupling, 
\begin{equation}
\frac{m_g}{T}
=
b_0 + b_1 \alpha + b_2 \alpha^2 + b_3 \alpha^3 ,
\qquad (T>T_c),
\label{eq:mg_highT}
\end{equation}
with
$b_0 = 0.218$,
$b_1 = 3.734$,
$b_2 = -1.160$,
and
$b_3 = 0.274$.
Here $\alpha \equiv \alpha_s/\pi = g^2/(4\pi^2)$ denotes the running strong coupling constant. For pure SU(3) gauge theory with $N_f=0$, the temperature dependence of the coupling is taken in the following renormalization-group-motivated form~\cite{Peng:2006tk}
\begin{equation}
\alpha
=
\frac{\beta_0}
{\beta_0^2 \ln(\Lambda/\Lambda_\text{QCD})
+ \beta_1 \ln\!\left[\ln(\Lambda/\Lambda_\text{QCD})\right]},
\label{eq:alpha}
\end{equation}
with $\beta_0 = 11/2$ and $\beta_1 = 51/4$.
The renormalization scale $\Lambda$ is chosen as a linear function of the reduced temperature,
$\Lambda/\Lambda_\text{QCD} = c_0 + c_1 x$,
with $c_0 = 1.054$ and $c_1 = 0.479$. 
This form describes the temperature evolution of the high-temperature branch from the strongly nonconformal region near $T_c$ to the high-temperature regime where approximate scale invariance is gradually restored. The use of scale-dependent couplings and effective masses has also been explored
in thermodynamically consistent descriptions of quark matter~\cite{Lu:2016fki,Ma:2023stj}.

\begin{figure}
\includegraphics[width=0.48\textwidth]{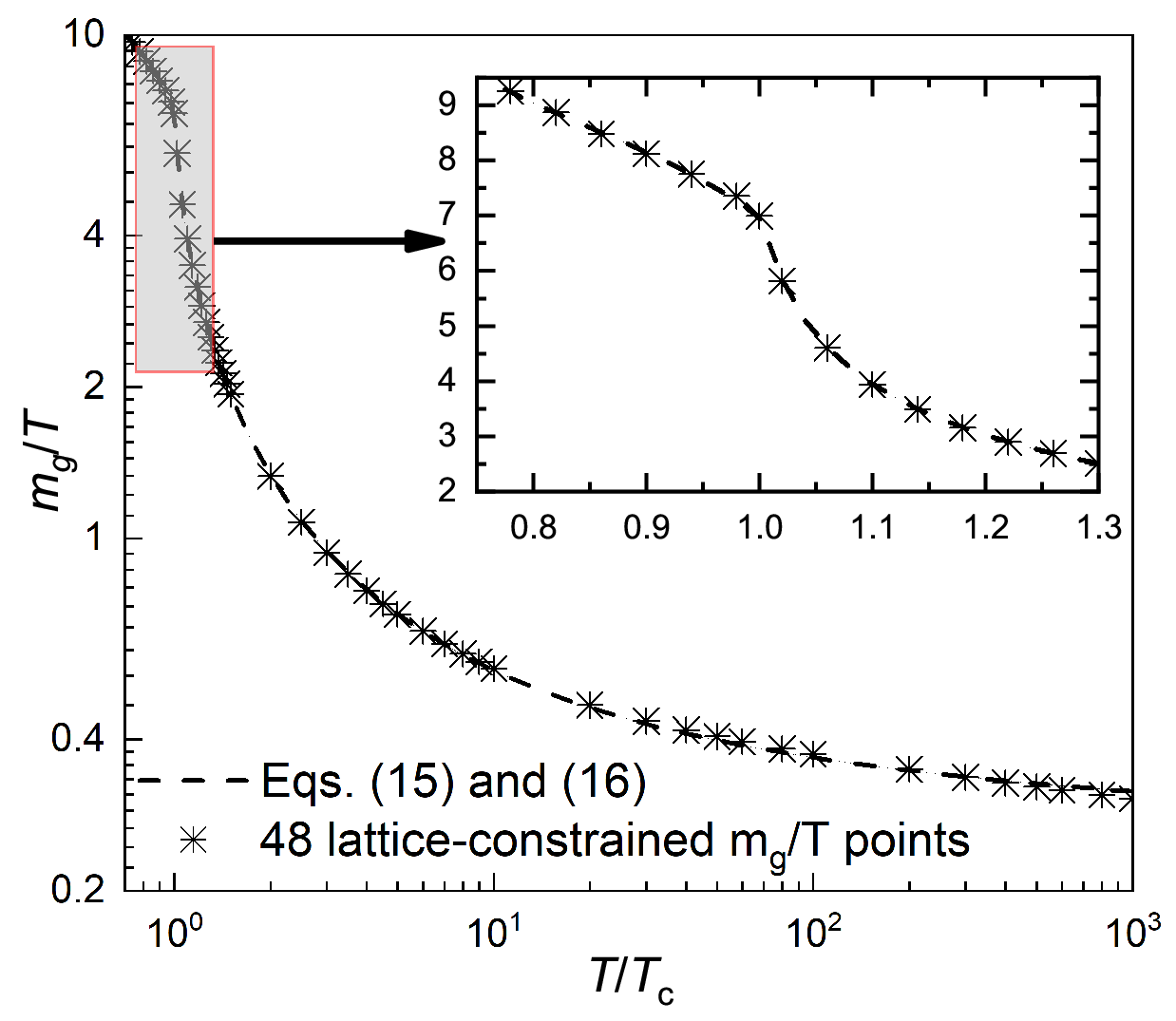}
\caption{
{ Temperature dependence of the effective gluon mass ratio $m_g/T$ as a function of the reduced temperature $T/T_c$. Symbols denote the pressure-extracted values obtained by matching the model pressure to the lattice pressure at each temperature point, while the solid line represents the fitted analytic parametrization in Eqs.~(\ref{eq:mg_lowT}) and (\ref{eq:mg_highT}). The inset shows an enlarged view of the near-$T_c$ region, where the effective mass changes rapidly.
}
\label{fig:mgT}
}
\end{figure}

The quality of the resulting parametrization is illustrated in
Fig.~\ref{fig:mgT}, where the pressure-extracted values of
$m_g/T$ are shown as functions of $T/T_c$. The fitted analytic
forms in Eqs.~(\ref{eq:mg_lowT}) and (\ref{eq:mg_highT}) reproduce
these discrete values very well over the whole temperature range
considered. The inset highlights the near-$T_c$ region, where
$m_g/T$ varies most rapidly and the parametrization remains in
good agreement with the pressure-extracted values.

We also clarify the matching prescription at the deconfinement temperature. The two branches of the effective mass profile are matched at the level of $m_g/T$ around $x=T/T_c=1$ within the accuracy of the lattice-constrained fit, so that no spurious discontinuity of the effective mass itself is introduced. However, no additional differentiability or smoothing condition is imposed across $T_c$. In particular, the derivatives of $m_g(T)$, and consequently derivative-sensitive thermodynamic quantities, are not forced to be continuous at the matching point. This treatment avoids artificially smoothing out the nonanalytic behavior expected near the first-order deconfinement transition of pure SU(3) gauge theory. The present parametrization should therefore be regarded as a lattice-constrained effective interpolation across the transition region, rather than as an attempt to resolve the exact singular structure at $T_c$.

Several remarks are in order. First, the effective mass introduced here should not be identified with the perturbative Debye screening mass, nor does it claim to reproduce the full gauge-invariant quasiparticle spectrum of thermal Yang-Mills theory. Rather, it is a phenomenological parameter entering the equilibrium thermodynamics. Second, below $T_c$ the use of the term ``effective gluon mass'' is merely a convenient notation for the parametrization of the lattice equation of state; it should not be interpreted as evidence for freely propagating gluons in the confined phase, where color-singlet excitations, in particular glueball-like degrees of freedom, are expected to dominate~\cite{Buisseret:2010mop,Trotti:2022knd}. Third, the nonvanishing asymptotic value implied by the present parametrization reflects the chosen lattice-constrained fit and should be regarded as an effective description over the temperature range of interest, rather than as a fundamental statement about the exact ultraviolet behavior of QCD. 
In practice, allowing for a small nonzero asymptotic value $b_0$ significantly improves the overall agreement with lattice thermodynamic data over a wide temperature range, especially in the intermediate and moderately high-temperature regions. 
Imposing $b_0=0$ would enforce the strict conformal limit at asymptotically high temperature, but leads to a noticeably less accurate description of the lattice equation of state. 
Consequently, the Stefan--Boltzmann values quoted below should be understood as massless conformal reference values, not as exact asymptotic limits enforced by the present fitted mass profile.

With the parametrization above and the thermodynamically consistent relations in Eqs.~(\ref{eq:entropy})-(\ref{eq:energy}), the TDM model provides a closed framework for evaluating not only the pressure, entropy density, and energy density, but also higher-order response functions. In this sense, the nontrivial content of the present analysis is not merely the reproduction of lattice-informed bulk observables, but the systematic prediction of derivative-sensitive quantities such as the specific heat and the isentropic bulk modulus, whose behavior provides a sharper diagnostic of nonconformal gluon dynamics.

It is useful to compare the present pressure-extracted mass profile with
representative quasiparticle descriptions in the literature. In standard
quasiparticle approaches, the effective gluon mass is usually introduced to
encode interaction effects in the deconfined medium, but its detailed
temperature dependence is model dependent. For example, quasiparticle
descriptions based on the trace anomaly, Polyakov-loop dynamics, or effective
gluon potentials can lead to different functional forms of $m_g/T$, even
when they are constrained by related lattice thermodynamic data
\cite{Plumari:2011mk,Sasaki:2012bi,Ruggieri:2012ny,Castorina:2011ja}.
In the present work, $m_g/T$ is instead obtained by directly matching the
model pressure to the lattice pressure at each temperature point and then
fitting the resulting pressure-extracted values. Therefore, the mass profile
used here should be regarded as an effective thermodynamic parametrization
specific to the present TDM framework, rather than as a universal
quasiparticle mass. A quantitative one-to-one comparison of $m_g/T$ between
different models should therefore be interpreted with care, since the mass
parameter is tied to the specific thermodynamic construction used in each
approach.

\section{Results and Discussion} \label{sec:results}

\subsection{Bulk thermodynamic quantities}

\begin{figure}
\includegraphics[width=0.48\textwidth]{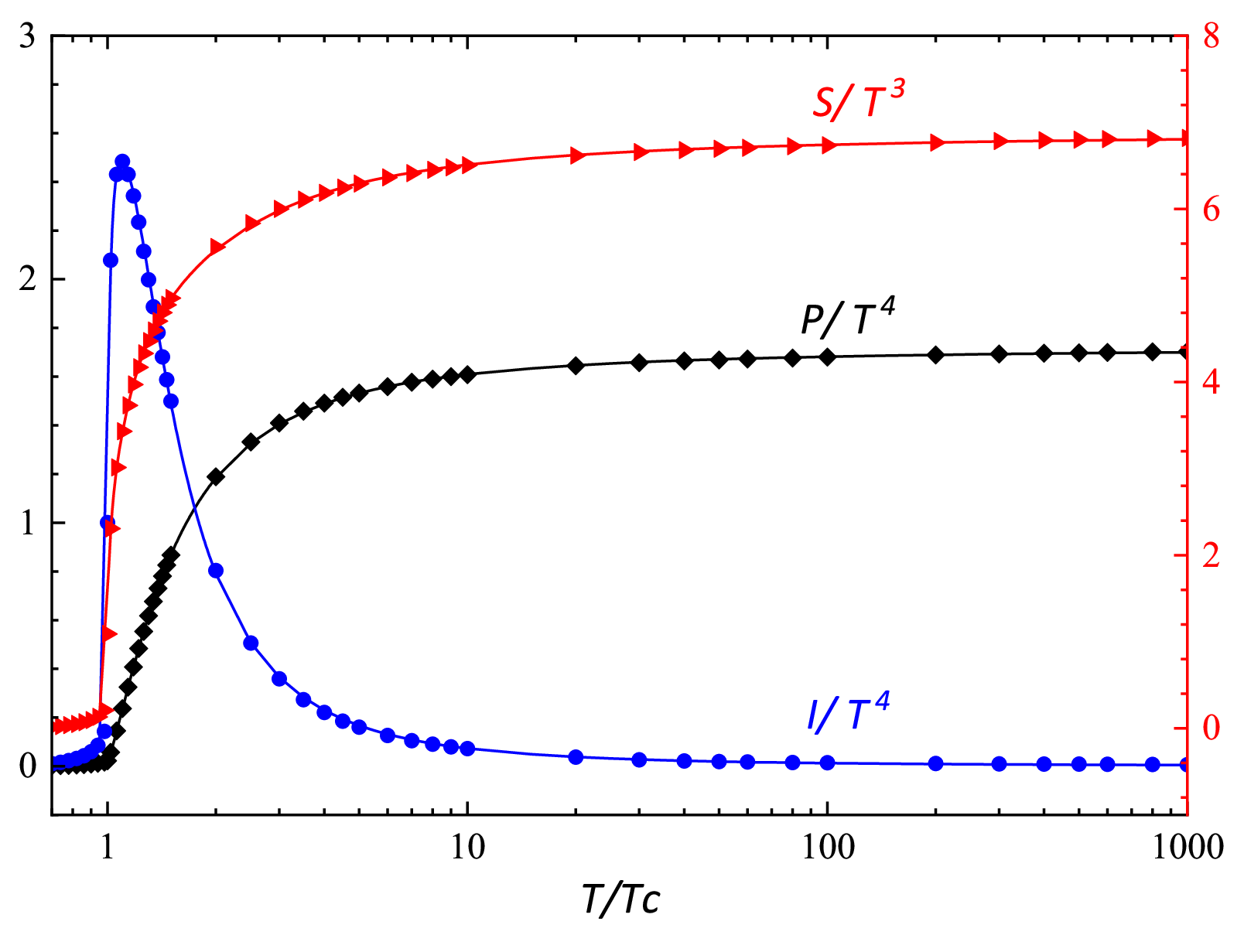}
\caption{
{Normalized pressure $P/T^4$ (black diamonds), trace anomaly $I/T^4$ (blue circles), and entropy density $S/T^3$ (red triangles) as functions of $T/T_c$ in pure SU(3) gauge theory. Symbols denote lattice data~\cite{Borsanyi:2012ve}, and solid lines are results from the present lattice-constrained model with a temperature-dependent effective gluon mass $m_g(T)$. Note that $S/T^3$ is shown with respect to the right-hand red $y$-axis.}
\label{fig:PIT4}
}
\end{figure}

Using the temperature-dependent effective gluon mass $m_g(T)$ as input, we
first evaluate the bulk thermodynamic observables of pure SU(3) gauge matter
within a thermodynamically consistent framework. These quantities have been
discussed in our previous work~\cite{Ma:2026xxx} and are revisited here to
establish a common baseline for the derivative thermodynamic response functions
analyzed below. 
{The normalized pressure $P/T^4$, trace anomaly $I/T^4$, and entropy density $S/T^3$} are
shown in Fig.~\ref{fig:PIT4} as functions of the reduced temperature $T/T_c$, together with lattice data for comparison. The trace anomaly, defined as $I \equiv E-3P$, provides a direct measure of the deviation from conformal behavior and is particularly sensitive to the temperature dependence of the underlying dynamics. {We emphasize that only the lattice pressure data are used to determine the effective mass profile; the trace anomaly and entropy density are subsequently calculated from the thermodynamically consistent relations and serve as consistency checks of the framework.}

As shown in Fig.~\ref{fig:PIT4}, all three observables are well described within the present framework across the deconfinement region. For clarity, $S/T^3$ is plotted on the right-hand red $y$-axis. The suppression of $P/T^4$ relative to the Stefan–Boltzmann limit, together with the pronounced enhancement of $I/T^4$, clearly reflects strong nonconformal behavior in the vicinity of $T_c$. In particular, the simultaneous description of $P/T^4$, $I/T^4$, and $S/T^3$ provides a nontrivial internal consistency check of the thermodynamic framework. All three quantities are obtained from the same pressure-constrained temperature-dependent effective gluon mass $m_g(T)$, rather than being independently fitted inputs. This indicates that the present model is capable of coherently capturing both bulk pressure-related observables and entropy-dominated thermodynamic degrees of freedom within a unified and self-consistent description.

Within the TDM model, these features are governed by the temperature
evolution of the effective mass scale. Near $T_c$, a relatively large effective
mass suppresses thermally accessible gluonic degrees of freedom and induces a
strong temperature dependence of the energy density, which is reflected in the
peak structure of the trace anomaly. As the temperature increases, the
effective mass contribution becomes less dominant relative to the thermal
scale, and the system progressively approaches the conformal relation
$E\simeq 3P$, with $I/T^4$ decreasing toward zero. Correspondingly, $P/T^4$ moves toward the massless Stefan--Boltzmann reference value from below, signaling the gradual restoration of approximate scale invariance.

The trace anomaly also provides a useful connection between the bulk equation
of state and the response functions discussed in the following subsections. Through the identity $I = T^5 \frac{d}{dT}\left(\frac{P}{T^4}\right)$, it directly probes the temperature derivative of the equation of state and therefore provides a stringent test of the framework.
In the present work, this baseline plays a crucial role, as higher-order response functions such as the specific heat and the isentropic bulk modulus involve additional temperature derivatives and are therefore more sensitive to the detailed structure of the equation of state.
In this sense, the agreement with lattice data is not a trivial consequence of inputting a temperature-dependent mass, but rather reflects the fact that the chosen parametrization yields a thermodynamically consistent description of both the magnitude and the temperature variation of the equation of state.
The bulk thermodynamic results presented here thus serve as a nontrivial foundation for the analysis of second-order response functions discussed below.


\begin{figure}
\includegraphics[width=0.49\textwidth]{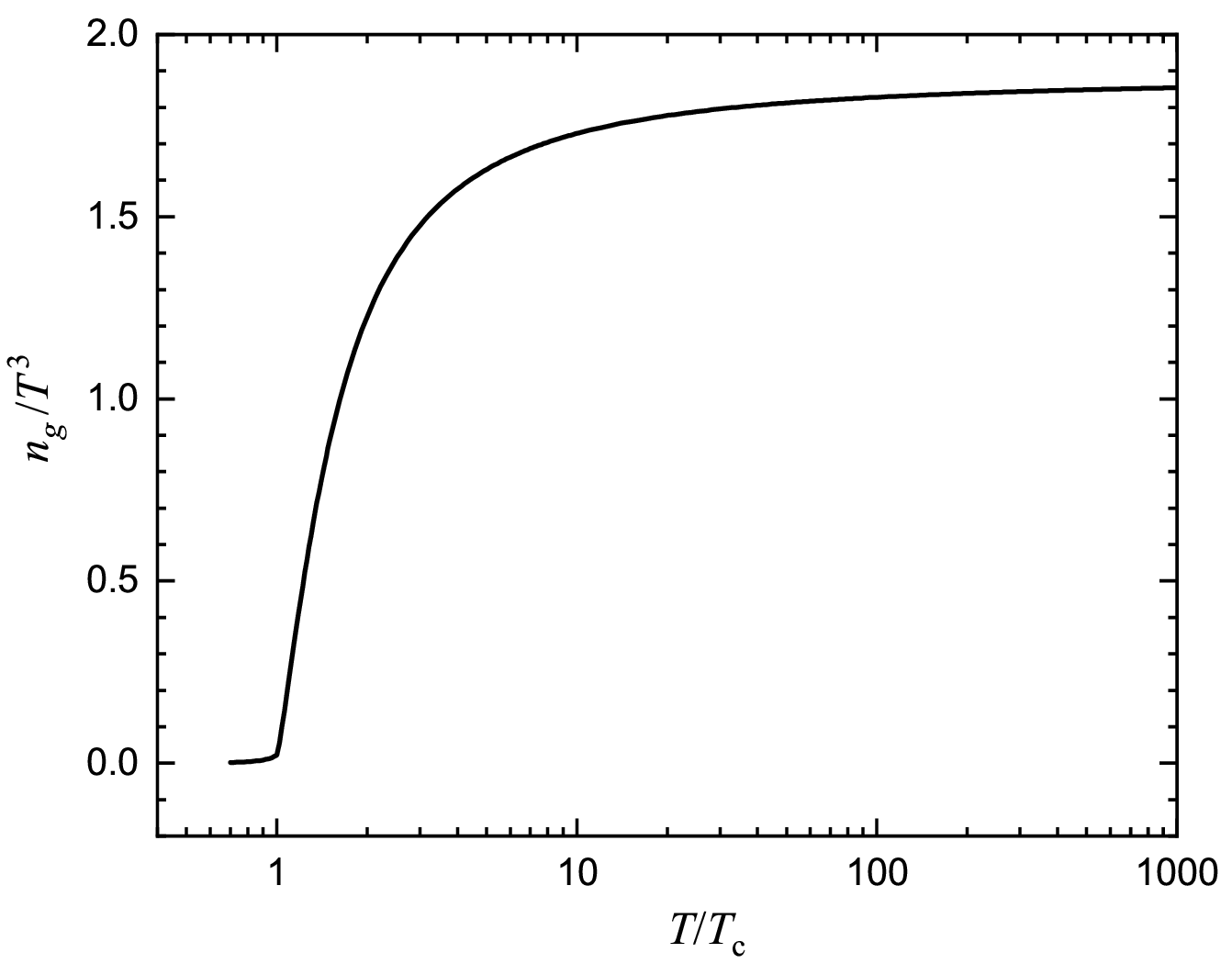}
\caption{Scaled gluon number density $n_g/T^3$ as a function of the reduced temperature $T/T_c$ in pure SU(3) gauge matter, obtained within the present TDM framework with a temperature-dependent effective gluon mass $m_g(T)$.
}\label{fig:nGT3}
\end{figure}

In Fig.~\ref{fig:nGT3}, we show the temperature dependence of the scaled gluon number density $n_g/T^3$ as a function of the reduced temperature $T/T_c$. Although $n_g$ is not associated with a conserved charge in pure gluon matter,
it provides a useful phenomenological measure of the thermally accessible
gluonic degrees of freedom within the present effective description. At temperatures below and near $T_c$, $n_g/T^3$ is strongly suppressed and remains close to zero. A similar suppression of effective gluonic degrees of freedom near the
deconfinement region also appears in Polyakov-loop-based phenomenological
descriptions of gluon thermodynamics~\cite{Meisinger:2001cq}. This suppression should be interpreted as an effective reduction of thermally accessible gluonic degrees of freedom, rather than the literal absence of excitations, reflecting the strongly interacting and nonconformal nature of the system in this regime. 
Above $T_c$, $n_g/T^3$ increases rapidly, indicating the activation of effective
gluonic degrees of freedom across the deconfinement region. As the temperature increases further, the growth gradually slows down and the ratio approaches
a high-temperature plateau. This tendency is consistent with the emergence of
ultrarelativistic scaling, for which the number density behaves approximately
as $n_g\propto T^3$ and the scaled quantity $n_g/T^3$ becomes nearly
temperature independent.

The behavior of $n_g/T^3$ is controlled mainly by the ratio $m_g/T$, which
determines the effective phase space available for thermal gluonic excitations.
Near $T_c$, the relatively large effective mass scale suppresses the thermal
occupation of gluonic modes and leads to the small value of $n_g/T^3$. As the
temperature increases, the effective mass contribution becomes less important
relative to the thermal scale, allowing a rapid growth of the thermal
population. At high temperatures, interaction effects become progressively
weaker compared with the dominant thermal scale, and the system approaches an
approximately scale-invariant regime. The rapid increase of $n_g/T^3$ near $T_c$ is closely correlated with the strong
temperature variation of the energy density and trace anomaly discussed above.
It therefore provides an intuitive picture of how effective gluonic degrees of
freedom are activated across the deconfinement region. This evolution also
sets the stage for the behavior of the derivative thermodynamic response
functions discussed below, in particular the enhancement of the specific
heat and the rapid change of the isentropic bulk modulus~\cite{Ruggieri:2012ny,Sasaki:2013xfa,Lo:2013hla}.

\begin{figure}
\includegraphics[width=0.49\textwidth]{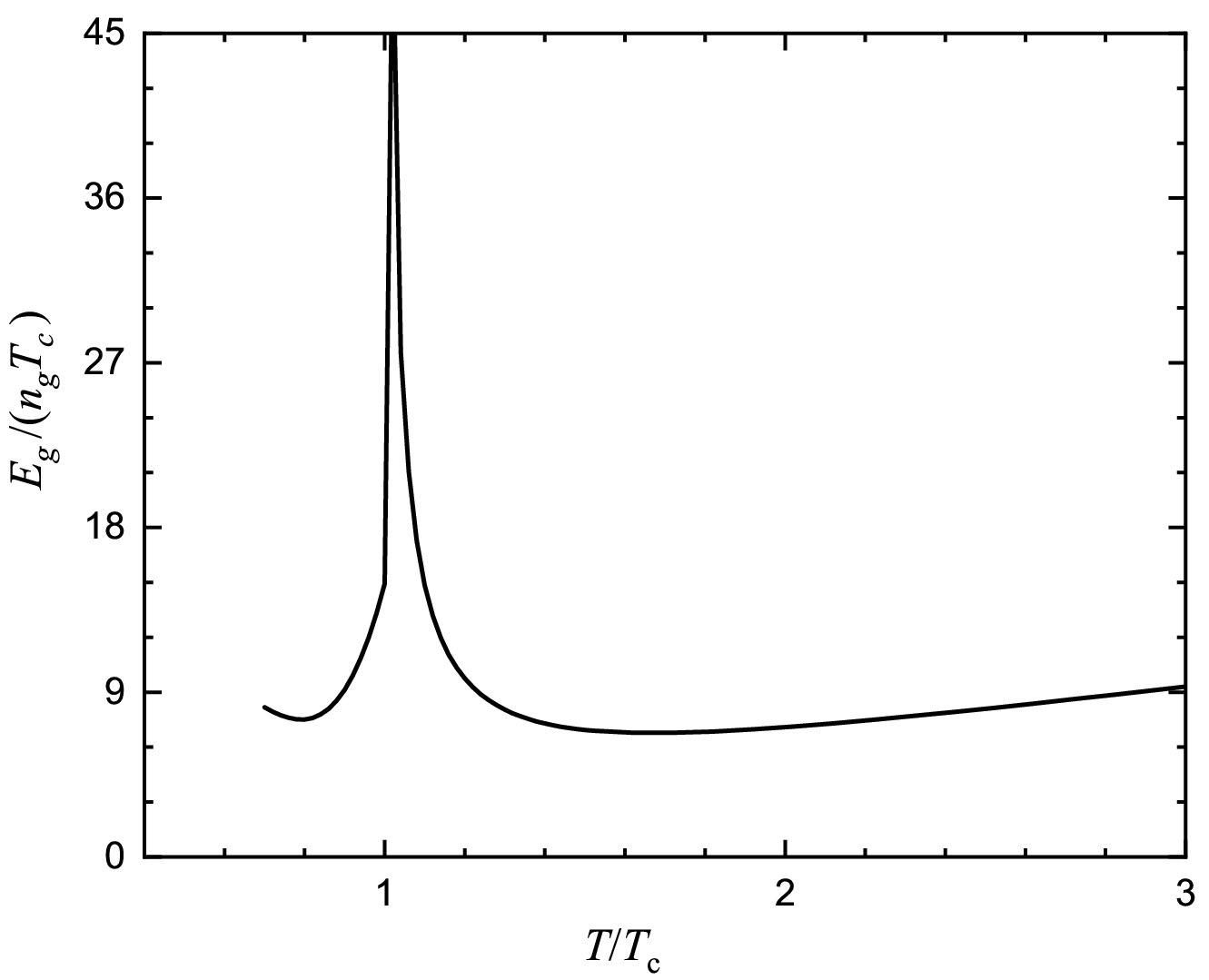}
\caption{Energy per thermally active gluonic mode normalized by $T_c$, $E_g/(n_g T_c)$, as a function of $T/T_c$ in pure SU(3) gluon matter within the present phenomenological framework. 
}\label{fig:EGnG}
\end{figure}

In Fig.~\ref{fig:EGnG}, we show the temperature dependence of the dimensionless ratio $E_g/(n_g T_c)$ as a function of $T/T_c$, where $E_g$ and $n_g$ denote the gluonic energy density and number density, respectively. This ratio characterizes the typical energy scale per thermally excited gluonic mode and provides complementary information to $n_g/T^3$. A pronounced nonmonotonic structure is observed in the vicinity of $T_c$: just above the transition, the ratio exhibits a sharp peak, followed by a rapid decrease to a local minimum at $T \gtrsim T_c$, and then a gradual increase at higher temperatures. The enhancement near $T_c$ originates from the
strong suppression of the number density relative to the energy density. In the
presence of a sizable effective mass $m_g(T)$, the thermal occupation of gluonic
modes is significantly reduced, such that $n_g$ is more strongly suppressed than
$E_g$. As a result, near $T_c$ the system is characterized by a relatively small
number of thermally excited modes that carry comparatively large energy,
leading to the peak structure in $E_g/(n_g T_c)$.

As the temperature increases above $T_c$, the ratio $m_g/T$ decreases and thermal
excitations are rapidly activated. In this regime, the number density grows more
quickly than the energy density, giving rise to the observed minimum in
$E_g/(n_g T_c)$. This behavior reflects a transition from a sparse population of
high-energy modes to a denser ensemble of moderately energetic excitations. At higher temperatures, the system gradually approaches an ultrarelativistic
regime characterized by $E_g \propto T^4$ and $n_g \propto T^3$, leading to
$E_g/(n_g T_c) \propto T/T_c$. The slow increase of the ratio at large $T/T_c$
is therefore consistent with dimensional scaling and signals the restoration of
approximate scale invariance.  Together with the behavior of $n_g/T^3$ in Fig.~\ref{fig:nGT3}, these results
provide a coherent thermodynamic picture across the deconfinement region. In
particular, the rapid variation of both quantities near $T_c$ reflects the
strong temperature dependence of the energy density, which underlies the pronounced enhancement of the specific heat discussed in the following subsection.

\subsection{Specific heat and isentropic bulk modulus}

\begin{figure}
\includegraphics[width=0.49\textwidth]{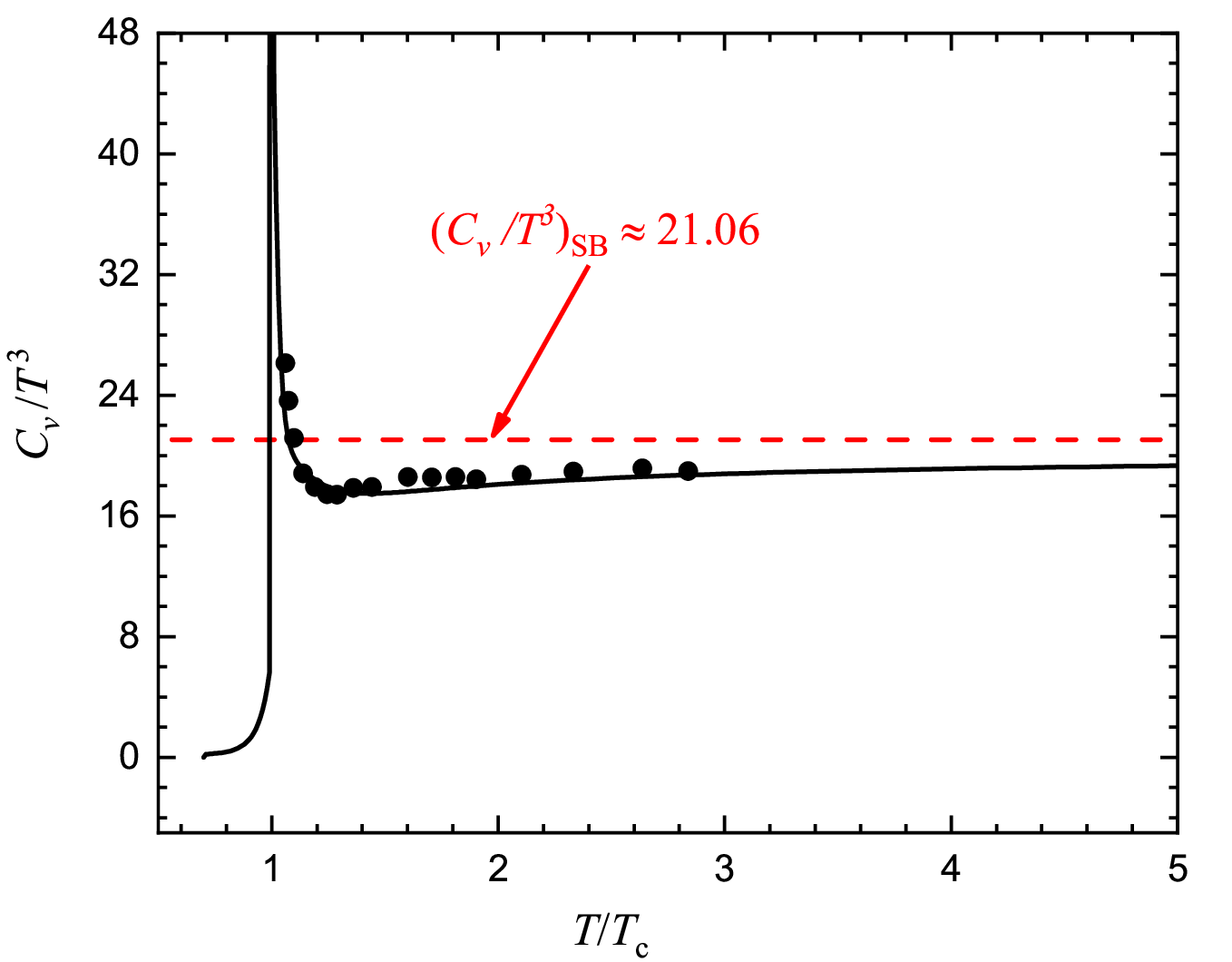}
\caption{Normalized specific heat $C_V/T^3$ as a function of the reduced temperature $T/T_c$. The dashed line indicates the massless Stefan--Boltzmann reference value $(C_V/T^3)_{\rm SB}=32\pi^2/15\simeq 21.06$. Lattice data (black circles) are taken from Refs.~\cite{Boyd:1996bx,Singha:2021wpi} and provide an independent comparison.
}\label{fig:CvT3}
\end{figure}

A more sensitive probe of nonconformal dynamics near the deconfinement temperature is provided by the specific heat, which measures the response of the energy density to temperature variations. Unlike bulk thermodynamic quantities such as the pressure or trace anomaly, which involve first-order derivatives, the specific heat corresponds to a second-order derivative of the thermodynamic potential and is therefore particularly sensitive to rapid changes in the equation of state. The specific heat at constant volume is given by~\cite{Gavai:2004se} 
\begin{equation}
\frac{C_V}{T^3}=\frac{1}{T^3}\frac{\partial E}{\partial T}\bigg|_V 
=4 \frac{E}{T^4}
+T \frac{\partial}{\partial T}\left(\frac{E}{T^4}\right) ,
\label{eq:Cv}
\end{equation}
which shows explicitly that deviations of $C_V/T^3$ from a constant are governed by the temperature dependence of $E/T^4$, and hence by violations of conformal scaling.


As shown in Fig.~\ref{fig:CvT3}, $C_V/T^3$ exhibits a pronounced near-$T_c$ enhancement. At low temperatures, it remains strongly suppressed,
reflecting the limited thermal excitation of effective degrees of freedom. As
the temperature approaches the deconfinement region, $C_V/T^3$ rises rapidly
and reaches a sharp maximum in the vicinity of $T_c$, followed by a sharp decrease to a
local minimum. At higher temperatures, it gradually increases again and tends
toward a constant value. This nonmonotonic behavior directly reflects the rapid
temperature variation of the energy density across the transition region. The pronounced enhancement originates from the strong temperature dependence of
$E/T^4$, which is closely correlated with the peak of the trace anomaly
$I/T^4$ and the rapid increase of the gluon number density shown in
Figs.~\ref{fig:PIT4} and \ref{fig:nGT3}. According to Eq.~(\ref{eq:Cv}), the
derivative term $T\,d(E/T^4)/dT$ amplifies this rapid variation, leading to a
pronounced enhancement of the specific heat. In this sense, $C_V$ provides a
more direct measure of the rate of change of the equation of state than bulk
quantities themselves.

The available lattice data for $C_V/T^3$ exhibit a similar qualitative behavior,
with an enhancement near $T_c$ followed by a gradual relaxation at higher
temperatures. The present result reproduces the overall trend and magnitude of
these lattice estimates above $T_c$, while noticeable differences appear in the
immediate vicinity of the peak. We emphasize that the lattice data for
$C_V/T^3$ shown here are not used as input in constraining the model, but rather
serve as an independent comparison. Since $C_V$ involves temperature derivatives
of the energy density, it is intrinsically more sensitive to the detailed
structure of the underlying equation of state, as well as to interpolation
procedures and finite-volume effects near the first-order deconfinement
transition. The observed deviations are therefore within the expected level of
sensitivity for such derivative-dependent quantities.

We have verified that reducing the temperature step size in the vicinity of $T_c$ further sharpens the structure of $C_V/T^3$, while leaving the behavior away from the critical region unchanged within numerical accuracy. This confirms that the observed peak-like feature is robust against discretization effects and is not an artifact of the numerical grid or the branch assignment at $x=1$.  
At sufficiently high temperatures, the system approaches an ultrarelativistic
regime characterized by $E \propto T^4$ and $C_V \propto T^3$, such that the
ratio $C_V/T^3$ asymptotically approaches a constant. The high-temperature behavior observed in Fig.~\ref{fig:CvT3} can be compared with the massless Stefan--Boltzmann reference value,
\begin{align}
\left(\frac{C_V}{T^3}\right)_{\rm SB}=\frac{32\pi^2}{15} \simeq 21.06.
\end{align}
This value follows from Eq.~(\ref{eq:Cv}) in the conformal massless limit, where $E\propto T^4$. In the present parametrization, however, $m_g(T)/T$ approaches the small nonzero value $b_0=0.218$ at asymptotically high temperature. Therefore, the Stefan--Boltzmann result should be interpreted as a conformal reference value rather than an exact asymptote of the fitted TDM model. This behavior indicates a gradual approach toward an approximately scale-invariant regime, accompanied by the progressive suppression of interaction effects. 
Because $C_V$ involves an additional temperature derivative, it provides a more stringent and complementary probe of the temperature dependence encoded in the effective gluon mass. 
Its nontrivial structure therefore reflects derivative-level features of the thermodynamics, rather than being determined solely by the bulk thermodynamic quantities themselves.

\begin{figure}
\includegraphics[width=0.49\textwidth]{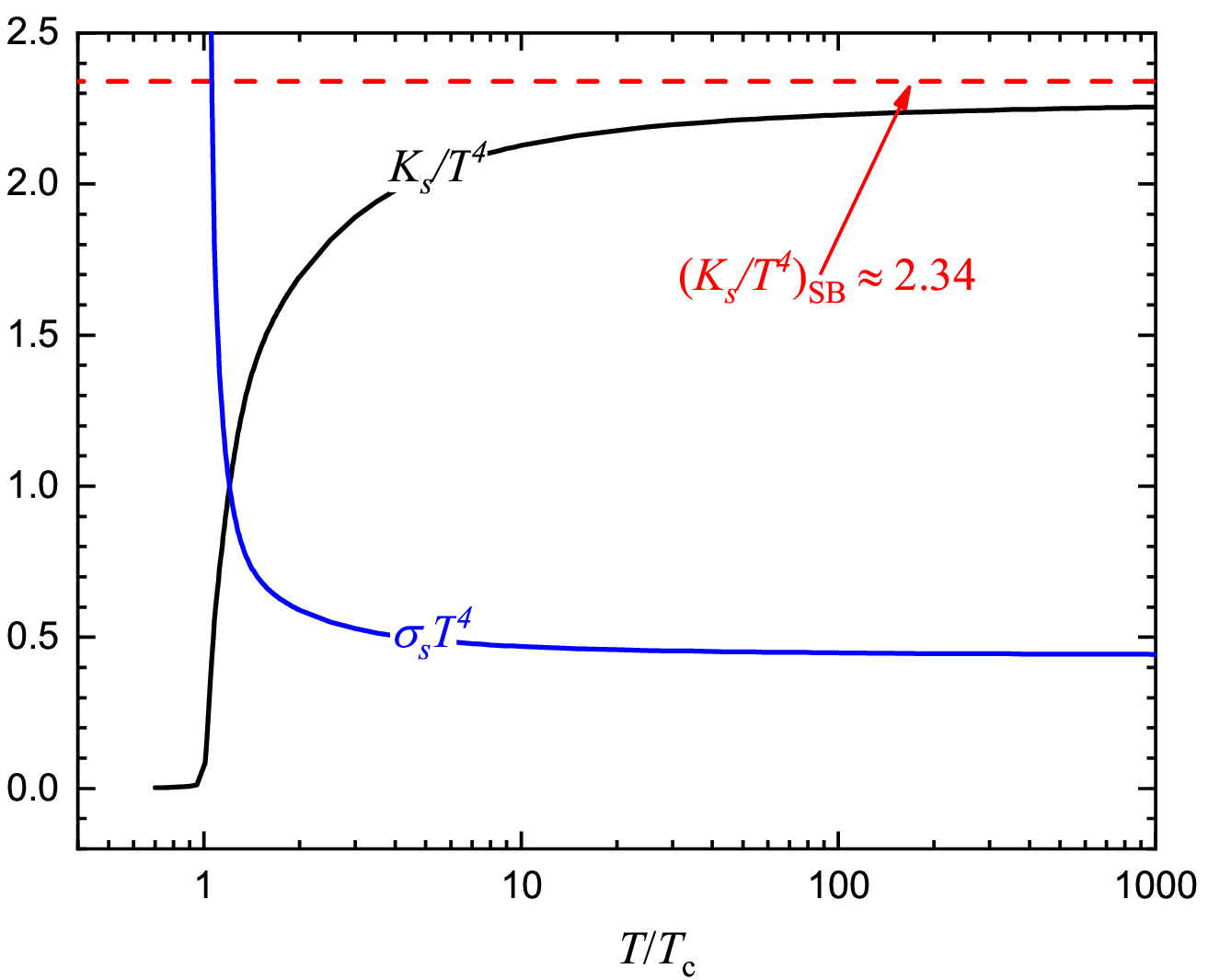}
\caption{Normalized isentropic bulk modulus $K_S/T^4$ and the corresponding dimensionless isentropic compressibility $\sigma_S T^4$ as functions of the reduced temperature $T/T_c$. The dashed horizontal line indicates the massless Stefan--Boltzmann reference value $(K_S/T^4)_{\rm SB}=32\pi^2/135\simeq 2.34$.
}\label{fig:betaKs}
\end{figure}

Finally, we turn to the adiabatic bulk modulus, which quantifies the mechanical stiffness of the medium under isentropic compression. For a relativistic system at vanishing baryon chemical potential, the bulk modulus is defined as
\begin{equation}
K_S= -V\left(\frac{\partial P}{\partial V}\right)_{S,\,N_B}
= (E+P) c_s^2,
\label{eq:KS_def}
\end{equation}
where $c_s^2=dP/dE$ is the squared speed of sound. The role of the speed of sound in governing the mechanical response and isentropic evolution of hot QCD matter has been emphasized in effective-model studies of heavy-ion dynamics~\cite{Motta:2020cbr}. This relation makes explicit that the bulk modulus is governed by both the enthalpy density and the propagation of sound modes, linking thermodynamic and mechanical properties of the system. Since $K_S$ is directly controlled by the speed of sound through $K_S=(E+P)c_s^2$, previous studies of the speed of sound in QCD and SU(3) Yang--Mills matter provide useful benchmarks for the mechanical response of the medium~\cite{Gavai:2004se,Khaidukov:2018lor}. It is also useful to note that the corresponding isentropic compressibility is given by $\sigma_S = 1/K_S$, providing an equivalent characterization of the mechanical response in terms of the ease of compression.

The temperature dependence of $K_S/T^4$ is shown in Fig.~\ref{fig:betaKs}. At low temperatures, $K_S/T^4$ is strongly suppressed, reflecting the limited thermodynamic response of the confined system. In this regime, both the pressure and the enthalpy density are small, and the medium offers little resistance to compression. As the temperature approaches the deconfinement region, $K_S/T^4$ increases rapidly, signaling a pronounced stiffening of the equation of state. This behavior originates from the combined effect of a rapidly increasing enthalpy density and the recovery of a finite speed of sound. Since $c_s^2$ measures how efficiently pressure builds up in response to changes in energy density, its increase indicates that the system becomes more effective at converting energy into mechanical pressure.

At higher temperatures, $K_S/T^4$ gradually approaches a constant plateau, signaling the onset of approximate scale invariance. In the massless conformal limit, where $E=3P$ and $c_s^2=1/3$, one obtains the reference value
\begin{equation}
\left(\frac{K_S}{T^4}\right)_{\rm SB}
=
\frac{32\pi^2}{135}
\simeq 2.34.
\end{equation}
Because the present fitted mass profile gives $m_g(T)/T\to b_0\neq0$, this Stefan--Boltzmann value is used here as a conformal reference rather than as an exact asymptotic limit of the model. 
The high-temperature behavior observed in Fig.~\ref{fig:betaKs} moves toward this conformal reference value, reflecting the progressive suppression of interaction effects. For completeness, the corresponding isentropic compressibility
$\sigma_S = 1/K_S$ is also shown in Fig.~\ref{fig:betaKs}. It exhibits the
opposite trend, decreasing rapidly across the deconfinement region and
approaching a constant value at high temperatures. In the massless conformal limit, the corresponding reference value is~\cite{Bollweg:2022fqq}
$
\left(\sigma_S T^4\right)_{\rm SB}
=
135/(32\pi^2)
\simeq 0.427$.  
This inverse behavior provides a complementary perspective on the mechanical
response: while $K_S$ characterizes stiffness, $\sigma_S$ quantifies the ease
of compression. Together with the behaviors of $n_g/T^3$ and $C_V/T^3$, the bulk modulus
provides an additional probe of the equation of state. While the specific heat
$C_V$ characterizes the thermal response through energy fluctuations, the
isentropic bulk modulus $K_S$ reflects the mechanical stiffness of the system.
Their combined behavior demonstrates how both thermodynamic and mechanical
responses evolve from a strongly interacting, nonconformal regime near $T_c$
to an approximately scale-invariant plasma at high temperature. Because $K_S$ depends on both the equation of state and its derivatives through
$c_s^2$, it provides an independent constraint on the temperature dependence
encoded in the effective gluon mass within the TDM framework.

\section{Conclusions}  \label{sec:CONCLUSION}

In this work, we have employed a lattice-data-constrained phenomenological
framework, referred to as the TDM model, to investigate the thermodynamic and
response properties of pure SU(3) gluon matter at finite temperature. Building
on our previous formulation, the present study extends the framework to
thermodynamic response functions and provides a more detailed characterization
of nonconformal dynamics. 
The bulk thermodynamic quantities, including the pressure, entropy density, and trace anomaly,
are well reproduced over a wide temperature range, indicating that the essential
features of the equation of state are captured. In particular, the pronounced
peak of the trace anomaly near $T_c$ reflects strong violations of conformal
scaling and encodes the underlying nonperturbative dynamics. Complementary information from the gluon number density and the energy per thermally active gluonic mode further illustrates how both the effective number of thermally active degrees of freedom and the characteristic energy scale evolve across the deconfinement region.  
The thermodynamic response functions provide a more sensitive probe of these
effects. The specific heat exhibits a pronounced enhancement near $T_c$, while the
isentropic bulk modulus increases rapidly across the transition, indicating a
strong stiffening of the equation of state. At higher temperatures, both $C_V/T^3$ and $K_S/T^4$ approach nearly constant values close to their massless Stefan--Boltzmann reference values, consistent with the gradual restoration of approximate scale invariance.

Because these response functions involve additional temperature derivatives of the thermodynamic potential, they provide a more stringent and complementary probe of the temperature dependence encoded in the effective gluon mass. Their nontrivial behavior reflects derivative-level features of the thermodynamics and cannot be inferred from the magnitudes of bulk thermodynamic quantities alone. 
Taken together, these results demonstrate that the TDM model provides a
thermodynamically consistent and physically transparent description of both bulk
and response properties of pure gauge matter. In particular, the framework
captures the interplay between effective degrees of freedom, energy scales, and
nonconformal dynamics across the deconfinement region. 
The present approach thus offers a minimal yet predictive baseline for studying derivative-sensitive observables and can be naturally extended to investigate transport properties and more realistic QCD systems at finite temperature~\cite{Rath:2022oum,Frasca:2025ffn, Plumari:2012ep}.

\section*{Acknowledgments}

We thank Cheng-Jun Xia for useful discussions.
This work is supported by  
the National Natural Science Foundation of China 
(Grants No.~12205093, No.~12405054, and No.~12375045), the Natural Science Foundation of Hunan Province (Grants No.~2026JJ50352, No.~2026JJ60316, and No.~2024JJ6210), and the Scientific Research Foundation of Hunan Provincial Education Department (Grants No.~25B0432 and No.~25C0302).

\bibliographystyle{aapmrev4-2}  
\bibliography{Ref}

\begin{thebibliography}{91}%
\makeatletter
\providecommand \@ifxundefined [1]{%
 \@ifx{#1\undefined}
}%
\providecommand \@ifnum [1]{%
 \ifnum #1\expandafter \@firstoftwo
 \else \expandafter \@secondoftwo
 \fi
}%
\providecommand \@ifx [1]{%
 \ifx #1\expandafter \@firstoftwo
 \else \expandafter \@secondoftwo
 \fi
}%
\providecommand \natexlab [1]{#1}%
\providecommand \enquote  [1]{``#1''}%
\providecommand \bibnamefont  [1]{#1}%
\providecommand \bibfnamefont [1]{#1}%
\providecommand \citenamefont [1]{#1}%
\providecommand \href@noop [0]{\@secondoftwo}%
\providecommand \href [0]{\begingroup \@sanitize@url \@href}%
\providecommand \@href[1]{\@@startlink{#1}\@@href}%
\providecommand \@@href[1]{\endgroup#1\@@endlink}%
\providecommand \@sanitize@url [0]{\catcode `\\12\catcode `\$12\catcode
  `\&12\catcode `\#12\catcode `\^12\catcode `\_12\catcode `\%12\relax}%
\providecommand \@@startlink[1]{}%
\providecommand \@@endlink[0]{}%
\providecommand \url  [0]{\begingroup\@sanitize@url \@url }%
\providecommand \@url [1]{\endgroup\@href {#1}{\urlprefix }}%
\providecommand \urlprefix  [0]{URL }%
\providecommand \Eprint [0]{\href }%
\providecommand \doibase [0]{https://doi.org/}%
\providecommand \selectlanguage [0]{\@gobble}%
\providecommand \bibinfo  [0]{\@secondoftwo}%
\providecommand \bibfield  [0]{\@secondoftwo}%
\providecommand \translation [1]{[#1]}%
\providecommand \BibitemOpen [0]{}%
\providecommand \bibitemStop [0]{}%
\providecommand \bibitemNoStop [0]{.\EOS\space}%
\providecommand \EOS [0]{\spacefactor3000\relax}%
\providecommand \BibitemShut  [1]{\csname bibitem#1\endcsname}%
\let\auto@bib@innerbib\@empty
\bibitem [{\citenamefont {Gross}\ \emph {et~al.}(2023)\citenamefont {Gross}
  \emph {et~al.}}]{Gross:2022hyw}%
  \BibitemOpen
  \bibfield  {author} {\bibinfo {author} {\bibfnamefont {F.}~\bibnamefont
  {Gross}} \emph {et~al.},\ }\href
  {https://doi.org/10.1140/epjc/s10052-023-11949-2} {\bibfield  {journal}
  {\bibinfo  {journal} {Eur. Phys. J. C}\ }\textbf {\bibinfo {volume} {83}},\
  \bibinfo {pages} {1125} (\bibinfo {year} {2023})},\ \Eprint
  {https://arxiv.org/abs/2212.11107} {arXiv:2212.11107 [hep-ph]} \BibitemShut
  {NoStop}%
\bibitem [{\citenamefont {Brambilla}\ \emph {et~al.}(2014)\citenamefont
  {Brambilla} \emph {et~al.}}]{Brambilla:2014jmp}%
  \BibitemOpen
  \bibfield  {author} {\bibinfo {author} {\bibfnamefont {N.}~\bibnamefont
  {Brambilla}} \emph {et~al.},\ }\href
  {https://doi.org/10.1140/epjc/s10052-014-2981-5} {\bibfield  {journal}
  {\bibinfo  {journal} {Eur. Phys. J. C}\ }\textbf {\bibinfo {volume} {74}},\
  \bibinfo {pages} {2981} (\bibinfo {year} {2014})},\ \Eprint
  {https://arxiv.org/abs/1404.3723} {arXiv:1404.3723 [hep-ph]} \BibitemShut
  {NoStop}%
\bibitem [{\citenamefont {Kumar}\ \emph {et~al.}(2024)\citenamefont {Kumar}
  \emph {et~al.}}]{MUSES:2023hyz}%
  \BibitemOpen
  \bibfield  {author} {\bibinfo {author} {\bibfnamefont {R.}~\bibnamefont
  {Kumar}} \emph {et~al.},\ }\href {https://doi.org/10.1007/s41114-024-00049-6}
  {\bibfield  {journal} {\bibinfo  {journal} {Living Rev. Rel.}\ }\textbf
  {\bibinfo {volume} {27}},\ \bibinfo {pages} {3} (\bibinfo {year}
  {2024})}\BibitemShut {NoStop}%
\bibitem [{\citenamefont {Adhikari}\ \emph {et~al.}(2026)\citenamefont
  {Adhikari} \emph {et~al.}}]{Adhikari:2024bfa}%
  \BibitemOpen
  \bibfield  {author} {\bibinfo {author} {\bibfnamefont {P.}~\bibnamefont
  {Adhikari}} \emph {et~al.},\ }\href
  {https://doi.org/10.1016/j.ppnp.2025.104199} {\bibfield  {journal} {\bibinfo
  {journal} {Prog. Part. Nucl. Phys.}\ }\textbf {\bibinfo {volume} {146}},\
  \bibinfo {pages} {104199} (\bibinfo {year} {2026})}\BibitemShut {NoStop}%
\bibitem [{\citenamefont {Bresciani}\ \emph {et~al.}(2026)\citenamefont
  {Bresciani}, \citenamefont {Brida}, \citenamefont {Giusti},\ and\
  \citenamefont {Pepe}}]{Bresciani:2025mcu}%
  \BibitemOpen
  \bibfield  {author} {\bibinfo {author} {\bibfnamefont {M.}~\bibnamefont
  {Bresciani}}, \bibinfo {author} {\bibfnamefont {M.~D.}\ \bibnamefont
  {Brida}}, \bibinfo {author} {\bibfnamefont {L.}~\bibnamefont {Giusti}},\ and\
  \bibinfo {author} {\bibfnamefont {M.}~\bibnamefont {Pepe}},\ }\href
  {https://doi.org/10.1103/jl9n-lk9k} {\bibfield  {journal} {\bibinfo
  {journal} {Phys. Rev. D}\ }\textbf {\bibinfo {volume} {113}},\ \bibinfo
  {pages} {034506} (\bibinfo {year} {2026})}\BibitemShut {NoStop}%
\bibitem [{\citenamefont {Chen}\ \emph {et~al.}(2024)\citenamefont {Chen} \emph
  {et~al.}}]{Chen:2024aom}%
  \BibitemOpen
  \bibfield  {author} {\bibinfo {author} {\bibfnamefont {J.}~\bibnamefont
  {Chen}} \emph {et~al.},\ }\href {https://doi.org/10.1007/s41365-024-01591-2}
  {\bibfield  {journal} {\bibinfo  {journal} {Nucl. Sci. Tech.}\ }\textbf
  {\bibinfo {volume} {35}},\ \bibinfo {pages} {214} (\bibinfo {year}
  {2024})}\BibitemShut {NoStop}%
\bibitem [{\citenamefont {Gyulassy}\ and\ \citenamefont
  {McLerran}(2005)}]{Gyulassy:2004zy}%
  \BibitemOpen
  \bibfield  {author} {\bibinfo {author} {\bibfnamefont {M.}~\bibnamefont
  {Gyulassy}}\ and\ \bibinfo {author} {\bibfnamefont {L.}~\bibnamefont
  {McLerran}},\ }\href {https://doi.org/10.1016/j.nuclphysa.2004.10.034}
  {\bibfield  {journal} {\bibinfo  {journal} {Nucl. Phys. A}\ }\textbf
  {\bibinfo {volume} {750}},\ \bibinfo {pages} {30} (\bibinfo {year} {2005})},\
  \Eprint {https://arxiv.org/abs/nucl-th/0405013} {arXiv:nucl-th/0405013}
  \BibitemShut {NoStop}%
\bibitem [{\citenamefont {Adcox}\ \emph {et~al.}(2005)\citenamefont {Adcox}
  \emph {et~al.}}]{PHENIX:2004vcz}%
  \BibitemOpen
  \bibfield  {author} {\bibinfo {author} {\bibfnamefont {K.}~\bibnamefont
  {Adcox}} \emph {et~al.},\ }\href
  {https://doi.org/10.1016/j.nuclphysa.2005.03.086} {\bibfield  {journal}
  {\bibinfo  {journal} {Nucl. Phys. A}\ }\textbf {\bibinfo {volume} {757}},\
  \bibinfo {pages} {184} (\bibinfo {year} {2005})}\BibitemShut {NoStop}%
\bibitem [{\citenamefont {Adams}\ \emph {et~al.}(2005)\citenamefont {Adams}
  \emph {et~al.}}]{STAR:2005gfr}%
  \BibitemOpen
  \bibfield  {author} {\bibinfo {author} {\bibfnamefont {J.}~\bibnamefont
  {Adams}} \emph {et~al.},\ }\href
  {https://doi.org/10.1016/j.nuclphysa.2005.03.085} {\bibfield  {journal}
  {\bibinfo  {journal} {Nucl. Phys. A}\ }\textbf {\bibinfo {volume} {757}},\
  \bibinfo {pages} {102} (\bibinfo {year} {2005})}\BibitemShut {NoStop}%
\bibitem [{\citenamefont {Arsene}\ \emph {et~al.}(2005)\citenamefont {Arsene}
  \emph {et~al.}}]{BRAHMS:2004adc}%
  \BibitemOpen
  \bibfield  {author} {\bibinfo {author} {\bibfnamefont {I.}~\bibnamefont
  {Arsene}} \emph {et~al.},\ }\href
  {https://doi.org/10.1016/j.nuclphysa.2005.02.130} {\bibfield  {journal}
  {\bibinfo  {journal} {Nucl. Phys. A}\ }\textbf {\bibinfo {volume} {757}},\
  \bibinfo {pages} {1} (\bibinfo {year} {2005})}\BibitemShut {NoStop}%
\bibitem [{\citenamefont {Shuryak}(2005)}]{Shuryak:2004cy}%
  \BibitemOpen
  \bibfield  {author} {\bibinfo {author} {\bibfnamefont {E.~V.}\ \bibnamefont
  {Shuryak}},\ }\href {https://doi.org/10.1016/j.nuclphysa.2004.10.022}
  {\bibfield  {journal} {\bibinfo  {journal} {Nucl. Phys. A}\ }\textbf
  {\bibinfo {volume} {750}},\ \bibinfo {pages} {64} (\bibinfo {year}
  {2005})}\BibitemShut {NoStop}%
\bibitem [{\citenamefont {Harris}\ and\ \citenamefont
  {M{\"u}ller}(2024)}]{Harris:2023tti}%
  \BibitemOpen
  \bibfield  {author} {\bibinfo {author} {\bibfnamefont {J.~W.}\ \bibnamefont
  {Harris}}\ and\ \bibinfo {author} {\bibfnamefont {B.}~\bibnamefont
  {M{\"u}ller}},\ }\href {https://doi.org/10.1140/epjc/s10052-024-12533-y}
  {\bibfield  {journal} {\bibinfo  {journal} {Eur. Phys. J. C}\ }\textbf
  {\bibinfo {volume} {84}},\ \bibinfo {pages} {247} (\bibinfo {year}
  {2024})}\BibitemShut {NoStop}%
\bibitem [{\citenamefont {Heinz}\ and\ \citenamefont
  {Snellings}(2013)}]{Heinz:2013th}%
  \BibitemOpen
  \bibfield  {author} {\bibinfo {author} {\bibfnamefont {U.}~\bibnamefont
  {Heinz}}\ and\ \bibinfo {author} {\bibfnamefont {R.}~\bibnamefont
  {Snellings}},\ }\href {https://doi.org/10.1146/annurev-nucl-102212-170540}
  {\bibfield  {journal} {\bibinfo  {journal} {Ann. Rev. Nucl. Part. Sci.}\
  }\textbf {\bibinfo {volume} {63}},\ \bibinfo {pages} {123} (\bibinfo {year}
  {2013})}\BibitemShut {NoStop}%
\bibitem [{\citenamefont {Shou}\ \emph {et~al.}(2024)\citenamefont {Shou} \emph
  {et~al.}}]{Shou:2024uga}%
  \BibitemOpen
  \bibfield  {author} {\bibinfo {author} {\bibfnamefont {Q.-Y.}\ \bibnamefont
  {Shou}} \emph {et~al.},\ }\href {https://doi.org/10.1007/s41365-024-01583-2}
  {\bibfield  {journal} {\bibinfo  {journal} {Nucl. Sci. Tech.}\ }\textbf
  {\bibinfo {volume} {35}},\ \bibinfo {pages} {219} (\bibinfo {year}
  {2024})}\BibitemShut {NoStop}%
\bibitem [{\citenamefont {Aarts}\ \emph {et~al.}(2023)\citenamefont {Aarts}
  \emph {et~al.}}]{Aarts:2023vsf}%
  \BibitemOpen
  \bibfield  {author} {\bibinfo {author} {\bibfnamefont {G.}~\bibnamefont
  {Aarts}} \emph {et~al.},\ }\href {https://doi.org/10.1016/j.ppnp.2023.104070}
  {\bibfield  {journal} {\bibinfo  {journal} {Prog. Part. Nucl. Phys.}\
  }\textbf {\bibinfo {volume} {133}},\ \bibinfo {pages} {104070} (\bibinfo
  {year} {2023})}\BibitemShut {NoStop}%
\bibitem [{\citenamefont {Acharya}\ \emph {et~al.}(2023)\citenamefont {Acharya}
  \emph {et~al.}}]{ALICE:2022hor}%
  \BibitemOpen
  \bibfield  {author} {\bibinfo {author} {\bibfnamefont {S.}~\bibnamefont
  {Acharya}} \emph {et~al.},\ }\href
  {https://doi.org/10.1103/PhysRevC.107.054617} {\bibfield  {journal} {\bibinfo
   {journal} {Phys. Rev. C}\ }\textbf {\bibinfo {volume} {107}},\ \bibinfo
  {pages} {054617} (\bibinfo {year} {2023})}\BibitemShut {NoStop}%
\bibitem [{\citenamefont {Ratti}(2018)}]{Ratti:2018ksb}%
  \BibitemOpen
  \bibfield  {author} {\bibinfo {author} {\bibfnamefont {C.}~\bibnamefont
  {Ratti}},\ }\href {https://doi.org/10.1088/1361-6633/aabb97} {\bibfield
  {journal} {\bibinfo  {journal} {Rept. Prog. Phys.}\ }\textbf {\bibinfo
  {volume} {81}},\ \bibinfo {pages} {084301} (\bibinfo {year} {2018})},\
  \Eprint {https://arxiv.org/abs/1804.07810} {arXiv:1804.07810 [hep-lat]}
  \BibitemShut {NoStop}%
\bibitem [{\citenamefont {Chen}\ \emph {et~al.}(2025)\citenamefont {Chen},
  \citenamefont {Chen}, \citenamefont {Li}, \citenamefont {Zhu},\ and\
  \citenamefont {Zhou}}]{Chen:2024epd}%
  \BibitemOpen
  \bibfield  {author} {\bibinfo {author} {\bibfnamefont {B.}~\bibnamefont
  {Chen}}, \bibinfo {author} {\bibfnamefont {X.}~\bibnamefont {Chen}}, \bibinfo
  {author} {\bibfnamefont {X.}~\bibnamefont {Li}}, \bibinfo {author}
  {\bibfnamefont {Z.-R.}\ \bibnamefont {Zhu}},\ and\ \bibinfo {author}
  {\bibfnamefont {K.}~\bibnamefont {Zhou}},\ }\href
  {https://doi.org/10.1103/PhysRevD.111.086033} {\bibfield  {journal} {\bibinfo
   {journal} {Phys. Rev. D}\ }\textbf {\bibinfo {volume} {111}},\ \bibinfo
  {pages} {086033} (\bibinfo {year} {2025})}\BibitemShut {NoStop}%
\bibitem [{\citenamefont {Altmann}\ \emph {et~al.}(2024)\citenamefont {Altmann}
  \emph {et~al.}}]{Altmann:2024icx}%
  \BibitemOpen
  \bibfield  {author} {\bibinfo {author} {\bibfnamefont {J.}~\bibnamefont
  {Altmann}} \emph {et~al.},\ }\href
  {https://doi.org/10.1140/epjc/s10052-024-12650-8} {\bibfield  {journal}
  {\bibinfo  {journal} {Eur. Phys. J. C}\ }\textbf {\bibinfo {volume} {84}},\
  \bibinfo {pages} {421} (\bibinfo {year} {2024})}\BibitemShut {NoStop}%
\bibitem [{\citenamefont {He}\ \emph {et~al.}(2026)\citenamefont {He},
  \citenamefont {Shao}, \citenamefont {Xie},\ and\ \citenamefont
  {Xu}}]{He:2024amc}%
  \BibitemOpen
  \bibfield  {author} {\bibinfo {author} {\bibfnamefont {W.-b.}\ \bibnamefont
  {He}}, \bibinfo {author} {\bibfnamefont {G.-y.}\ \bibnamefont {Shao}},
  \bibinfo {author} {\bibfnamefont {C.-l.}\ \bibnamefont {Xie}},\ and\ \bibinfo
  {author} {\bibfnamefont {R.-x.}\ \bibnamefont {Xu}},\ }\href
  {https://doi.org/10.1103/fp2p-rkpp} {\bibfield  {journal} {\bibinfo
  {journal} {Phys. Rev. D}\ }\textbf {\bibinfo {volume} {113}},\ \bibinfo
  {pages} {014036} (\bibinfo {year} {2026})}\BibitemShut {NoStop}%
\bibitem [{\citenamefont {Ding}, \citenamefont {Shu},\ and\ \citenamefont
  {Zhang}(2026)}]{Ding:2026gco}%
  \BibitemOpen
  \bibfield  {author} {\bibinfo {author} {\bibfnamefont {H.-T.}\ \bibnamefont
  {Ding}}, \bibinfo {author} {\bibfnamefont {H.-T.}\ \bibnamefont {Shu}},\ and\
  \bibinfo {author} {\bibfnamefont {C.}~\bibnamefont {Zhang}},\ }\href
  {https://doi.org/10.1103/1h7r-clj7} {\bibfield  {journal} {\bibinfo
  {journal} {Phys. Rev. D}\ }\textbf {\bibinfo {volume} {113}},\ \bibinfo
  {pages} {074503} (\bibinfo {year} {2026})}\BibitemShut {NoStop}%
\bibitem [{\citenamefont {Astrakhantsev}\ \emph {et~al.}(2020)\citenamefont
  {Astrakhantsev}, \citenamefont {Braguta}, \citenamefont {D'Elia},
  \citenamefont {Kotov}, \citenamefont {Nikolaev},\ and\ \citenamefont
  {Sanfilippo}}]{Astrakhantsev:2019zkr}%
  \BibitemOpen
  \bibfield  {author} {\bibinfo {author} {\bibfnamefont {N.}~\bibnamefont
  {Astrakhantsev}}, \bibinfo {author} {\bibfnamefont {V.~V.}\ \bibnamefont
  {Braguta}}, \bibinfo {author} {\bibfnamefont {M.}~\bibnamefont {D'Elia}},
  \bibinfo {author} {\bibfnamefont {A.~{\relax Yu}.}\ \bibnamefont {Kotov}},
  \bibinfo {author} {\bibfnamefont {A.~A.}\ \bibnamefont {Nikolaev}},\ and\
  \bibinfo {author} {\bibfnamefont {F.}~\bibnamefont {Sanfilippo}},\ }\href
  {https://doi.org/10.1103/PhysRevD.102.054516} {\bibfield  {journal} {\bibinfo
   {journal} {Phys. Rev. D}\ }\textbf {\bibinfo {volume} {102}},\ \bibinfo
  {pages} {054516} (\bibinfo {year} {2020})},\ \Eprint
  {https://arxiv.org/abs/1910.08516} {arXiv:1910.08516 [hep-lat]} \BibitemShut
  {NoStop}%
\bibitem [{\citenamefont {Borsanyi}\ \emph {et~al.}(2018)\citenamefont
  {Borsanyi}, \citenamefont {Fodor}, \citenamefont {Guenther}, \citenamefont
  {Katz}, \citenamefont {Szabo}, \citenamefont {Pasztor}, \citenamefont
  {Portillo},\ and\ \citenamefont {Ratti}}]{Borsanyi:2018grb}%
  \BibitemOpen
  \bibfield  {author} {\bibinfo {author} {\bibfnamefont {S.}~\bibnamefont
  {Borsanyi}}, \bibinfo {author} {\bibfnamefont {Z.}~\bibnamefont {Fodor}},
  \bibinfo {author} {\bibfnamefont {J.~N.}\ \bibnamefont {Guenther}}, \bibinfo
  {author} {\bibfnamefont {S.~K.}\ \bibnamefont {Katz}}, \bibinfo {author}
  {\bibfnamefont {K.~K.}\ \bibnamefont {Szabo}}, \bibinfo {author}
  {\bibfnamefont {A.}~\bibnamefont {Pasztor}}, \bibinfo {author} {\bibfnamefont
  {I.}~\bibnamefont {Portillo}},\ and\ \bibinfo {author} {\bibfnamefont
  {C.}~\bibnamefont {Ratti}},\ }\href {https://doi.org/10.1007/JHEP10(2018)205}
  {\bibfield  {journal} {\bibinfo  {journal} {JHEP}\ }\textbf {\bibinfo
  {volume} {10}},\ \bibinfo {pages} {205} (\bibinfo {year} {2018})}\BibitemShut
  {NoStop}%
\bibitem [{\citenamefont {Puglisi}, \citenamefont {Plumari},\ and\
  \citenamefont {Greco}(2015)}]{Puglisi:2014pda}%
  \BibitemOpen
  \bibfield  {author} {\bibinfo {author} {\bibfnamefont {A.}~\bibnamefont
  {Puglisi}}, \bibinfo {author} {\bibfnamefont {S.}~\bibnamefont {Plumari}},\
  and\ \bibinfo {author} {\bibfnamefont {V.}~\bibnamefont {Greco}},\ }\href
  {https://doi.org/10.1016/j.physletb.2015.10.070} {\bibfield  {journal}
  {\bibinfo  {journal} {Phys. Lett. B}\ }\textbf {\bibinfo {volume} {751}},\
  \bibinfo {pages} {326} (\bibinfo {year} {2015})},\ \Eprint
  {https://arxiv.org/abs/1407.2559} {arXiv:1407.2559 [hep-ph]} \BibitemShut
  {NoStop}%
\bibitem [{\citenamefont {Zhao}\ \emph {et~al.}(2017)\citenamefont {Zhao},
  \citenamefont {Shi}, \citenamefont {Li},\ and\ \citenamefont
  {Zong}}]{Zhao:2017vmx}%
  \BibitemOpen
  \bibfield  {author} {\bibinfo {author} {\bibfnamefont {A.~M.}\ \bibnamefont
  {Zhao}}, \bibinfo {author} {\bibfnamefont {Y.-M.}\ \bibnamefont {Shi}},
  \bibinfo {author} {\bibfnamefont {J.-F.}\ \bibnamefont {Li}},\ and\ \bibinfo
  {author} {\bibfnamefont {H.-S.}\ \bibnamefont {Zong}},\ }\href
  {https://doi.org/10.1088/1674-1137/41/10/103101} {\bibfield  {journal}
  {\bibinfo  {journal} {Chin. Phys. C}\ }\textbf {\bibinfo {volume} {41}},\
  \bibinfo {pages} {103101} (\bibinfo {year} {2017})}\BibitemShut {NoStop}%
\bibitem [{\citenamefont {Parisi}\ \emph {et~al.}(2026)\citenamefont {Parisi},
  \citenamefont {Nugara}, \citenamefont {Plumari},\ and\ \citenamefont
  {Greco}}]{Parisi:2025gwq}%
  \BibitemOpen
  \bibfield  {author} {\bibinfo {author} {\bibfnamefont {G.}~\bibnamefont
  {Parisi}}, \bibinfo {author} {\bibfnamefont {V.}~\bibnamefont {Nugara}},
  \bibinfo {author} {\bibfnamefont {S.}~\bibnamefont {Plumari}},\ and\ \bibinfo
  {author} {\bibfnamefont {V.}~\bibnamefont {Greco}},\ }\href
  {https://doi.org/10.1103/qd2t-2sdp} {\bibfield  {journal} {\bibinfo
  {journal} {Phys. Rev. D}\ }\textbf {\bibinfo {volume} {113}},\ \bibinfo
  {pages} {014001} (\bibinfo {year} {2026})}\BibitemShut {NoStop}%
\bibitem [{\citenamefont {Bazavov}, \citenamefont {Petreczky},\ and\
  \citenamefont {Weber}(2018)}]{Bazavov:2017dsy}%
  \BibitemOpen
  \bibfield  {author} {\bibinfo {author} {\bibfnamefont {A.}~\bibnamefont
  {Bazavov}}, \bibinfo {author} {\bibfnamefont {P.}~\bibnamefont {Petreczky}},\
  and\ \bibinfo {author} {\bibfnamefont {J.~H.}\ \bibnamefont {Weber}},\ }\href
  {https://doi.org/10.1103/PhysRevD.97.014510} {\bibfield  {journal} {\bibinfo
  {journal} {Phys. Rev. D}\ }\textbf {\bibinfo {volume} {97}},\ \bibinfo
  {pages} {014510} (\bibinfo {year} {2018})},\ \Eprint
  {https://arxiv.org/abs/1710.05024} {arXiv:1710.05024 [hep-lat]} \BibitemShut
  {NoStop}%
\bibitem [{\citenamefont {Borsanyi}\ \emph {et~al.}(2014)\citenamefont
  {Borsanyi}, \citenamefont {Fodor}, \citenamefont {Hoelbling}, \citenamefont
  {Katz}, \citenamefont {Krieg},\ and\ \citenamefont
  {Szabo}}]{Borsanyi:2013bia}%
  \BibitemOpen
  \bibfield  {author} {\bibinfo {author} {\bibfnamefont {S.}~\bibnamefont
  {Borsanyi}}, \bibinfo {author} {\bibfnamefont {Z.}~\bibnamefont {Fodor}},
  \bibinfo {author} {\bibfnamefont {C.}~\bibnamefont {Hoelbling}}, \bibinfo
  {author} {\bibfnamefont {S.~D.}\ \bibnamefont {Katz}}, \bibinfo {author}
  {\bibfnamefont {S.}~\bibnamefont {Krieg}},\ and\ \bibinfo {author}
  {\bibfnamefont {K.~K.}\ \bibnamefont {Szabo}},\ }\href
  {https://doi.org/10.1016/j.physletb.2014.01.007} {\bibfield  {journal}
  {\bibinfo  {journal} {Phys. Lett. B}\ }\textbf {\bibinfo {volume} {730}},\
  \bibinfo {pages} {99} (\bibinfo {year} {2014})},\ \Eprint
  {https://arxiv.org/abs/1309.5258} {arXiv:1309.5258 [hep-lat]} \BibitemShut
  {NoStop}%
\bibitem [{\citenamefont {Aoki}\ \emph {et~al.}(2006)\citenamefont {Aoki},
  \citenamefont {Fodor}, \citenamefont {Katz},\ and\ \citenamefont
  {Szabo}}]{Aoki:2005vt}%
  \BibitemOpen
  \bibfield  {author} {\bibinfo {author} {\bibfnamefont {Y.}~\bibnamefont
  {Aoki}}, \bibinfo {author} {\bibfnamefont {Z.}~\bibnamefont {Fodor}},
  \bibinfo {author} {\bibfnamefont {S.~D.}\ \bibnamefont {Katz}},\ and\
  \bibinfo {author} {\bibfnamefont {K.~K.}\ \bibnamefont {Szabo}},\ }\href
  {https://doi.org/10.1088/1126-6708/2006/01/089} {\bibfield  {journal}
  {\bibinfo  {journal} {JHEP}\ }\textbf {\bibinfo {volume} {01}},\ \bibinfo
  {pages} {089} (\bibinfo {year} {2006})}\BibitemShut {NoStop}%
\bibitem [{\citenamefont {Borsanyi}\ \emph
  {et~al.}(2012{\natexlab{a}})\citenamefont {Borsanyi}, \citenamefont
  {Endrodi}, \citenamefont {Fodor}, \citenamefont {Katz}, \citenamefont
  {Krieg}, \citenamefont {Ratti},\ and\ \citenamefont
  {Szabo}}]{Borsanyi:2012cr}%
  \BibitemOpen
  \bibfield  {author} {\bibinfo {author} {\bibfnamefont {{\relax
  Sz}.}~\bibnamefont {Borsanyi}}, \bibinfo {author} {\bibfnamefont
  {G.}~\bibnamefont {Endrodi}}, \bibinfo {author} {\bibfnamefont
  {Z.}~\bibnamefont {Fodor}}, \bibinfo {author} {\bibfnamefont {S.~D.}\
  \bibnamefont {Katz}}, \bibinfo {author} {\bibfnamefont {S.}~\bibnamefont
  {Krieg}}, \bibinfo {author} {\bibfnamefont {C.}~\bibnamefont {Ratti}},\ and\
  \bibinfo {author} {\bibfnamefont {K.~K.}\ \bibnamefont {Szabo}},\ }\href
  {https://doi.org/10.1007/JHEP08(2012)053} {\bibfield  {journal} {\bibinfo
  {journal} {JHEP}\ }\textbf {\bibinfo {volume} {08}},\ \bibinfo {pages} {053}
  (\bibinfo {year} {2012}{\natexlab{a}})},\ \Eprint
  {https://arxiv.org/abs/1204.6710} {arXiv:1204.6710 [hep-lat]} \BibitemShut
  {NoStop}%
\bibitem [{\citenamefont {Borsanyi}\ \emph {et~al.}(2010)\citenamefont
  {Borsanyi}, \citenamefont {Endrodi}, \citenamefont {Fodor}, \citenamefont
  {Jakovac}, \citenamefont {Katz}, \citenamefont {Krieg}, \citenamefont
  {Ratti},\ and\ \citenamefont {Szabo}}]{Borsanyi:2010cj}%
  \BibitemOpen
  \bibfield  {author} {\bibinfo {author} {\bibfnamefont {S.}~\bibnamefont
  {Borsanyi}}, \bibinfo {author} {\bibfnamefont {G.}~\bibnamefont {Endrodi}},
  \bibinfo {author} {\bibfnamefont {Z.}~\bibnamefont {Fodor}}, \bibinfo
  {author} {\bibfnamefont {A.}~\bibnamefont {Jakovac}}, \bibinfo {author}
  {\bibfnamefont {S.~D.}\ \bibnamefont {Katz}}, \bibinfo {author}
  {\bibfnamefont {S.}~\bibnamefont {Krieg}}, \bibinfo {author} {\bibfnamefont
  {C.}~\bibnamefont {Ratti}},\ and\ \bibinfo {author} {\bibfnamefont {K.~K.}\
  \bibnamefont {Szabo}},\ }\href {https://doi.org/10.1007/JHEP11(2010)077}
  {\bibfield  {journal} {\bibinfo  {journal} {JHEP}\ }\textbf {\bibinfo
  {volume} {11}},\ \bibinfo {pages} {077} (\bibinfo {year} {2010})},\ \Eprint
  {https://arxiv.org/abs/1007.2580} {arXiv:1007.2580 [hep-lat]} \BibitemShut
  {NoStop}%
\bibitem [{\citenamefont {Caselle}, \citenamefont {Nada},\ and\ \citenamefont
  {Panero}(2018)}]{Caselle:2018kap}%
  \BibitemOpen
  \bibfield  {author} {\bibinfo {author} {\bibfnamefont {M.}~\bibnamefont
  {Caselle}}, \bibinfo {author} {\bibfnamefont {A.}~\bibnamefont {Nada}},\ and\
  \bibinfo {author} {\bibfnamefont {M.}~\bibnamefont {Panero}},\ }\href
  {https://doi.org/10.1103/PhysRevD.98.054513} {\bibfield  {journal} {\bibinfo
  {journal} {Phys. Rev. D}\ }\textbf {\bibinfo {volume} {98}},\ \bibinfo
  {pages} {054513} (\bibinfo {year} {2018})},\ \Eprint
  {https://arxiv.org/abs/1801.03110} {arXiv:1801.03110 [hep-lat]} \BibitemShut
  {NoStop}%
\bibitem [{\citenamefont {Borsanyi}\ \emph
  {et~al.}(2012{\natexlab{b}})\citenamefont {Borsanyi}, \citenamefont
  {Endrodi}, \citenamefont {Fodor}, \citenamefont {Katz},\ and\ \citenamefont
  {Szabo}}]{Borsanyi:2012ve}%
  \BibitemOpen
  \bibfield  {author} {\bibinfo {author} {\bibfnamefont {{\relax
  Sz}.}~\bibnamefont {Borsanyi}}, \bibinfo {author} {\bibfnamefont
  {G.}~\bibnamefont {Endrodi}}, \bibinfo {author} {\bibfnamefont
  {Z.}~\bibnamefont {Fodor}}, \bibinfo {author} {\bibfnamefont {S.~D.}\
  \bibnamefont {Katz}},\ and\ \bibinfo {author} {\bibfnamefont {K.~K.}\
  \bibnamefont {Szabo}},\ }\href {https://doi.org/10.1007/JHEP07(2012)056}
  {\bibfield  {journal} {\bibinfo  {journal} {JHEP}\ }\textbf {\bibinfo
  {volume} {07}},\ \bibinfo {pages} {056} (\bibinfo {year}
  {2012}{\natexlab{b}})},\ \Eprint {https://arxiv.org/abs/1204.6184}
  {arXiv:1204.6184 [hep-lat]} \BibitemShut {NoStop}%
\bibitem [{\citenamefont {Giusti}\ and\ \citenamefont
  {Pepe}(2017)}]{Giusti:2016iqr}%
  \BibitemOpen
  \bibfield  {author} {\bibinfo {author} {\bibfnamefont {L.}~\bibnamefont
  {Giusti}}\ and\ \bibinfo {author} {\bibfnamefont {M.}~\bibnamefont {Pepe}},\
  }\href {https://doi.org/10.1016/j.physletb.2017.04.001} {\bibfield  {journal}
  {\bibinfo  {journal} {Phys. Lett. B}\ }\textbf {\bibinfo {volume} {769}},\
  \bibinfo {pages} {385} (\bibinfo {year} {2017})}\BibitemShut {NoStop}%
\bibitem [{\citenamefont {Asakawa}\ \emph {et~al.}(2014)\citenamefont
  {Asakawa}, \citenamefont {Hatsuda}, \citenamefont {Itou}, \citenamefont
  {Kitazawa},\ and\ \citenamefont {Suzuki}}]{Asakawa:2013laa}%
  \BibitemOpen
  \bibfield  {author} {\bibinfo {author} {\bibfnamefont {M.}~\bibnamefont
  {Asakawa}}, \bibinfo {author} {\bibfnamefont {T.}~\bibnamefont {Hatsuda}},
  \bibinfo {author} {\bibfnamefont {E.}~\bibnamefont {Itou}}, \bibinfo {author}
  {\bibfnamefont {M.}~\bibnamefont {Kitazawa}},\ and\ \bibinfo {author}
  {\bibfnamefont {H.}~\bibnamefont {Suzuki}},\ }\href
  {https://doi.org/10.1103/PhysRevD.90.011501} {\bibfield  {journal} {\bibinfo
  {journal} {Phys. Rev. D}\ }\textbf {\bibinfo {volume} {90}},\ \bibinfo
  {pages} {011501} (\bibinfo {year} {2014})}\BibitemShut {NoStop}%
\bibitem [{\citenamefont {Boyd}\ \emph {et~al.}(1996)\citenamefont {Boyd},
  \citenamefont {Engels}, \citenamefont {Karsch}, \citenamefont {Laermann},
  \citenamefont {Legeland}, \citenamefont {L{\"u}tgemeier},\ and\ \citenamefont
  {Petersson}}]{Boyd:1996bx}%
  \BibitemOpen
  \bibfield  {author} {\bibinfo {author} {\bibfnamefont {G.}~\bibnamefont
  {Boyd}}, \bibinfo {author} {\bibfnamefont {J.}~\bibnamefont {Engels}},
  \bibinfo {author} {\bibfnamefont {F.}~\bibnamefont {Karsch}}, \bibinfo
  {author} {\bibfnamefont {E.}~\bibnamefont {Laermann}}, \bibinfo {author}
  {\bibfnamefont {C.}~\bibnamefont {Legeland}}, \bibinfo {author}
  {\bibfnamefont {M.}~\bibnamefont {L{\"u}tgemeier}},\ and\ \bibinfo {author}
  {\bibfnamefont {B.}~\bibnamefont {Petersson}},\ }\href
  {https://doi.org/10.1016/0550-3213(96)00170-8} {\bibfield  {journal}
  {\bibinfo  {journal} {Nucl. Phys. B}\ }\textbf {\bibinfo {volume} {469}},\
  \bibinfo {pages} {419} (\bibinfo {year} {1996})}\BibitemShut {NoStop}%
\bibitem [{\citenamefont {Giusti}\ \emph {et~al.}(2025)\citenamefont {Giusti},
  \citenamefont {Hirasawa}, \citenamefont {Pepe},\ and\ \citenamefont
  {Virz{\`i}}}]{Giusti:2025fxu}%
  \BibitemOpen
  \bibfield  {author} {\bibinfo {author} {\bibfnamefont {L.}~\bibnamefont
  {Giusti}}, \bibinfo {author} {\bibfnamefont {M.}~\bibnamefont {Hirasawa}},
  \bibinfo {author} {\bibfnamefont {M.}~\bibnamefont {Pepe}},\ and\ \bibinfo
  {author} {\bibfnamefont {L.}~\bibnamefont {Virz{\`i}}},\ }\href
  {https://doi.org/10.1016/j.physletb.2025.139775} {\bibfield  {journal}
  {\bibinfo  {journal} {Phys. Lett. B}\ }\textbf {\bibinfo {volume} {868}},\
  \bibinfo {pages} {139775} (\bibinfo {year} {2025})}\BibitemShut {NoStop}%
\bibitem [{\citenamefont {Plumari}\ \emph {et~al.}(2011)\citenamefont
  {Plumari}, \citenamefont {Alberico}, \citenamefont {Greco},\ and\
  \citenamefont {Ratti}}]{Plumari:2011mk}%
  \BibitemOpen
  \bibfield  {author} {\bibinfo {author} {\bibfnamefont {S.}~\bibnamefont
  {Plumari}}, \bibinfo {author} {\bibfnamefont {W.~M.}\ \bibnamefont
  {Alberico}}, \bibinfo {author} {\bibfnamefont {V.}~\bibnamefont {Greco}},\
  and\ \bibinfo {author} {\bibfnamefont {C.}~\bibnamefont {Ratti}},\ }\href
  {https://doi.org/10.1103/PhysRevD.84.094004} {\bibfield  {journal} {\bibinfo
  {journal} {Phys. Rev. D}\ }\textbf {\bibinfo {volume} {84}},\ \bibinfo
  {pages} {094004} (\bibinfo {year} {2011})},\ \Eprint
  {https://arxiv.org/abs/1103.5611} {arXiv:1103.5611 [hep-ph]} \BibitemShut
  {NoStop}%
\bibitem [{\citenamefont {Sasaki}\ and\ \citenamefont
  {Redlich}(2012)}]{Sasaki:2012bi}%
  \BibitemOpen
  \bibfield  {author} {\bibinfo {author} {\bibfnamefont {C.}~\bibnamefont
  {Sasaki}}\ and\ \bibinfo {author} {\bibfnamefont {K.}~\bibnamefont
  {Redlich}},\ }\href {https://doi.org/10.1103/PhysRevD.86.014007} {\bibfield
  {journal} {\bibinfo  {journal} {Phys. Rev. D}\ }\textbf {\bibinfo {volume}
  {86}},\ \bibinfo {pages} {014007} (\bibinfo {year} {2012})}\BibitemShut
  {NoStop}%
\bibitem [{\citenamefont {Ruggieri}\ \emph {et~al.}(2012)\citenamefont
  {Ruggieri}, \citenamefont {Alba}, \citenamefont {Castorina}, \citenamefont
  {Plumari}, \citenamefont {Ratti},\ and\ \citenamefont
  {Greco}}]{Ruggieri:2012ny}%
  \BibitemOpen
  \bibfield  {author} {\bibinfo {author} {\bibfnamefont {M.}~\bibnamefont
  {Ruggieri}}, \bibinfo {author} {\bibfnamefont {P.}~\bibnamefont {Alba}},
  \bibinfo {author} {\bibfnamefont {P.}~\bibnamefont {Castorina}}, \bibinfo
  {author} {\bibfnamefont {S.}~\bibnamefont {Plumari}}, \bibinfo {author}
  {\bibfnamefont {C.}~\bibnamefont {Ratti}},\ and\ \bibinfo {author}
  {\bibfnamefont {V.}~\bibnamefont {Greco}},\ }\href
  {https://doi.org/10.1103/PhysRevD.86.054007} {\bibfield  {journal} {\bibinfo
  {journal} {Phys. Rev. D}\ }\textbf {\bibinfo {volume} {86}},\ \bibinfo
  {pages} {054007} (\bibinfo {year} {2012})},\ \Eprint
  {https://arxiv.org/abs/1204.5995} {arXiv:1204.5995 [hep-ph]} \BibitemShut
  {NoStop}%
\bibitem [{\citenamefont {Castorina}, \citenamefont {Miller},\ and\
  \citenamefont {Satz}(2011)}]{Castorina:2011ja}%
  \BibitemOpen
  \bibfield  {author} {\bibinfo {author} {\bibfnamefont {P.}~\bibnamefont
  {Castorina}}, \bibinfo {author} {\bibfnamefont {D.~E.}\ \bibnamefont
  {Miller}},\ and\ \bibinfo {author} {\bibfnamefont {H.}~\bibnamefont {Satz}},\
  }\href {https://doi.org/10.1140/epjc/s10052-011-1673-7} {\bibfield  {journal}
  {\bibinfo  {journal} {Eur. Phys. J. C}\ }\textbf {\bibinfo {volume} {71}},\
  \bibinfo {pages} {1673} (\bibinfo {year} {2011})}\BibitemShut {NoStop}%
\bibitem [{\citenamefont {Heshmatian}\ and\ \citenamefont
  {Morad}(2024)}]{Heshmatian:2023yzz}%
  \BibitemOpen
  \bibfield  {author} {\bibinfo {author} {\bibfnamefont {S.}~\bibnamefont
  {Heshmatian}}\ and\ \bibinfo {author} {\bibfnamefont {R.}~\bibnamefont
  {Morad}},\ }\href {https://doi.org/10.1140/epjc/s10052-024-12596-x}
  {\bibfield  {journal} {\bibinfo  {journal} {Eur. Phys. J. C}\ }\textbf
  {\bibinfo {volume} {84}},\ \bibinfo {pages} {360} (\bibinfo {year}
  {2024})}\BibitemShut {NoStop}%
\bibitem [{\citenamefont {Chen}, \citenamefont {Giannuzzi},\ and\ \citenamefont
  {Nicotri}(2026)}]{Chen:2025goz}%
  \BibitemOpen
  \bibfield  {author} {\bibinfo {author} {\bibfnamefont {X.}~\bibnamefont
  {Chen}}, \bibinfo {author} {\bibfnamefont {F.}~\bibnamefont {Giannuzzi}},\
  and\ \bibinfo {author} {\bibfnamefont {S.}~\bibnamefont {Nicotri}},\ }\href
  {https://doi.org/10.1103/j7wh-vf6x} {\bibfield  {journal} {\bibinfo
  {journal} {Phys. Rev. D}\ }\textbf {\bibinfo {volume} {113}},\ \bibinfo
  {pages} {106007} (\bibinfo {year} {2026})}\BibitemShut {NoStop}%
\bibitem [{\citenamefont {He}\ \emph {et~al.}(2024)\citenamefont {He},
  \citenamefont {Li}, \citenamefont {Li},\ and\ \citenamefont
  {Wang}}]{He:2022amv}%
  \BibitemOpen
  \bibfield  {author} {\bibinfo {author} {\bibfnamefont {S.}~\bibnamefont
  {He}}, \bibinfo {author} {\bibfnamefont {L.}~\bibnamefont {Li}}, \bibinfo
  {author} {\bibfnamefont {Z.}~\bibnamefont {Li}},\ and\ \bibinfo {author}
  {\bibfnamefont {S.-J.}\ \bibnamefont {Wang}},\ }\href
  {https://doi.org/10.1007/s11433-023-2293-2} {\bibfield  {journal} {\bibinfo
  {journal} {Sci. China Phys. Mech. Astron.}\ }\textbf {\bibinfo {volume}
  {67}},\ \bibinfo {pages} {240411} (\bibinfo {year} {2024})}\BibitemShut
  {NoStop}%
\bibitem [{\citenamefont {Chen}, \citenamefont {Chen},\ and\ \citenamefont
  {Zhou}(2025)}]{Chen:2025kqb}%
  \BibitemOpen
  \bibfield  {author} {\bibinfo {author} {\bibfnamefont {X.}~\bibnamefont
  {Chen}}, \bibinfo {author} {\bibfnamefont {Y.}~\bibnamefont {Chen}},\ and\
  \bibinfo {author} {\bibfnamefont {K.}~\bibnamefont {Zhou}},\ }\href
  {https://doi.org/10.1103/wyr3-2kc5} {\bibfield  {journal} {\bibinfo
  {journal} {Phys. Rev. D}\ }\textbf {\bibinfo {volume} {112}},\ \bibinfo
  {pages} {126025} (\bibinfo {year} {2025})}\BibitemShut {NoStop}%
\bibitem [{\citenamefont {Yaresko}\ and\ \citenamefont
  {Kampfer}(2015)}]{Yaresko:2013tia}%
  \BibitemOpen
  \bibfield  {author} {\bibinfo {author} {\bibfnamefont {R.}~\bibnamefont
  {Yaresko}}\ and\ \bibinfo {author} {\bibfnamefont {B.}~\bibnamefont
  {Kampfer}},\ }\href {https://doi.org/10.1016/j.physletb.2015.05.034}
  {\bibfield  {journal} {\bibinfo  {journal} {Phys. Lett. B}\ }\textbf
  {\bibinfo {volume} {747}},\ \bibinfo {pages} {36} (\bibinfo {year}
  {2015})}\BibitemShut {NoStop}%
\bibitem [{\citenamefont {Alba}\ \emph {et~al.}(2014)\citenamefont {Alba},
  \citenamefont {Alberico}, \citenamefont {Bluhm}, \citenamefont {Greco},
  \citenamefont {Ratti},\ and\ \citenamefont {Ruggieri}}]{Alba:2014lda}%
  \BibitemOpen
  \bibfield  {author} {\bibinfo {author} {\bibfnamefont {P.}~\bibnamefont
  {Alba}}, \bibinfo {author} {\bibfnamefont {W.}~\bibnamefont {Alberico}},
  \bibinfo {author} {\bibfnamefont {M.}~\bibnamefont {Bluhm}}, \bibinfo
  {author} {\bibfnamefont {V.}~\bibnamefont {Greco}}, \bibinfo {author}
  {\bibfnamefont {C.}~\bibnamefont {Ratti}},\ and\ \bibinfo {author}
  {\bibfnamefont {M.}~\bibnamefont {Ruggieri}},\ }\href
  {https://doi.org/10.1016/j.nuclphysa.2014.11.011} {\bibfield  {journal}
  {\bibinfo  {journal} {Nucl. Phys. A}\ }\textbf {\bibinfo {volume} {934}},\
  \bibinfo {pages} {41} (\bibinfo {year} {2014})},\ \Eprint
  {https://arxiv.org/abs/1402.6213} {arXiv:1402.6213 [hep-ph]} \BibitemShut
  {NoStop}%
\bibitem [{\citenamefont {Ding}, \citenamefont {Karsch},\ and\ \citenamefont
  {Mukherjee}(2015)}]{Ding:2015ona}%
  \BibitemOpen
  \bibfield  {author} {\bibinfo {author} {\bibfnamefont {H.-T.}\ \bibnamefont
  {Ding}}, \bibinfo {author} {\bibfnamefont {F.}~\bibnamefont {Karsch}},\ and\
  \bibinfo {author} {\bibfnamefont {S.}~\bibnamefont {Mukherjee}},\ }\href
  {https://doi.org/10.1142/S0218301315300076} {\bibfield  {journal} {\bibinfo
  {journal} {Int. J. Mod. Phys. E}\ }\textbf {\bibinfo {volume} {24}},\
  \bibinfo {pages} {1530007} (\bibinfo {year} {2015})},\ \Eprint
  {https://arxiv.org/abs/1504.05274} {arXiv:1504.05274 [hep-lat]} \BibitemShut
  {NoStop}%
\bibitem [{\citenamefont {Bazavov}\ \emph {et~al.}(2014)\citenamefont {Bazavov}
  \emph {et~al.}}]{HotQCD:2014kol}%
  \BibitemOpen
  \bibfield  {author} {\bibinfo {author} {\bibfnamefont {A.}~\bibnamefont
  {Bazavov}} \emph {et~al.} (\bibinfo {collaboration} {HotQCD}),\ }\href
  {https://doi.org/10.1103/PhysRevD.90.094503} {\bibfield  {journal} {\bibinfo
  {journal} {Phys. Rev. D}\ }\textbf {\bibinfo {volume} {90}},\ \bibinfo
  {pages} {094503} (\bibinfo {year} {2014})},\ \Eprint
  {https://arxiv.org/abs/1407.6387} {arXiv:1407.6387 [hep-lat]} \BibitemShut
  {NoStop}%
\bibitem [{\citenamefont {Lu}\ \emph {et~al.}(2026)\citenamefont {Lu},
  \citenamefont {Gao}, \citenamefont {Liu},\ and\ \citenamefont
  {Pawlowski}}]{Lu:2025cls}%
  \BibitemOpen
  \bibfield  {author} {\bibinfo {author} {\bibfnamefont {Y.}~\bibnamefont
  {Lu}}, \bibinfo {author} {\bibfnamefont {F.}~\bibnamefont {Gao}}, \bibinfo
  {author} {\bibfnamefont {Y.-x.}\ \bibnamefont {Liu}},\ and\ \bibinfo {author}
  {\bibfnamefont {J.~M.}\ \bibnamefont {Pawlowski}},\ }\href
  {https://doi.org/10.1103/s1kg-cl9p} {\bibfield  {journal} {\bibinfo
  {journal} {Phys. Rev. D}\ }\textbf {\bibinfo {volume} {113}},\ \bibinfo
  {pages} {054019} (\bibinfo {year} {2026})}\BibitemShut {NoStop}%
\bibitem [{\citenamefont {Fu}\ \emph {et~al.}(2016)\citenamefont {Fu},
  \citenamefont {Pawlowski}, \citenamefont {Rennecke},\ and\ \citenamefont
  {Schaefer}}]{Fu:2016tey}%
  \BibitemOpen
  \bibfield  {author} {\bibinfo {author} {\bibfnamefont {W.-j.}\ \bibnamefont
  {Fu}}, \bibinfo {author} {\bibfnamefont {J.~M.}\ \bibnamefont {Pawlowski}},
  \bibinfo {author} {\bibfnamefont {F.}~\bibnamefont {Rennecke}},\ and\
  \bibinfo {author} {\bibfnamefont {B.-J.}\ \bibnamefont {Schaefer}},\ }\href
  {https://doi.org/10.1103/PhysRevD.94.116020} {\bibfield  {journal} {\bibinfo
  {journal} {Phys. Rev. D}\ }\textbf {\bibinfo {volume} {94}},\ \bibinfo
  {pages} {116020} (\bibinfo {year} {2016})},\ \Eprint
  {https://arxiv.org/abs/1608.04302} {arXiv:1608.04302 [hep-ph]} \BibitemShut
  {NoStop}%
\bibitem [{\citenamefont {{Braun-Munzinger}}\ \emph {et~al.}(2021)\citenamefont
  {{Braun-Munzinger}}, \citenamefont {Friman}, \citenamefont {Redlich},
  \citenamefont {Rustamov},\ and\ \citenamefont
  {Stachel}}]{Braun-Munzinger:2020jbk}%
  \BibitemOpen
  \bibfield  {author} {\bibinfo {author} {\bibfnamefont {P.}~\bibnamefont
  {{Braun-Munzinger}}}, \bibinfo {author} {\bibfnamefont {B.}~\bibnamefont
  {Friman}}, \bibinfo {author} {\bibfnamefont {K.}~\bibnamefont {Redlich}},
  \bibinfo {author} {\bibfnamefont {A.}~\bibnamefont {Rustamov}},\ and\
  \bibinfo {author} {\bibfnamefont {J.}~\bibnamefont {Stachel}},\ }\href
  {https://doi.org/10.1016/j.nuclphysa.2021.122141} {\bibfield  {journal}
  {\bibinfo  {journal} {Nucl. Phys. A}\ }\textbf {\bibinfo {volume} {1008}},\
  \bibinfo {pages} {122141} (\bibinfo {year} {2021})}\BibitemShut {NoStop}%
\bibitem [{\citenamefont {Sorensen}\ \emph {et~al.}(2021)\citenamefont
  {Sorensen}, \citenamefont {Oliinychenko}, \citenamefont {Koch},\ and\
  \citenamefont {McLerran}}]{Sorensen:2021zme}%
  \BibitemOpen
  \bibfield  {author} {\bibinfo {author} {\bibfnamefont {A.}~\bibnamefont
  {Sorensen}}, \bibinfo {author} {\bibfnamefont {D.}~\bibnamefont
  {Oliinychenko}}, \bibinfo {author} {\bibfnamefont {V.}~\bibnamefont {Koch}},\
  and\ \bibinfo {author} {\bibfnamefont {L.}~\bibnamefont {McLerran}},\ }\href
  {https://doi.org/10.1103/PhysRevLett.127.042303} {\bibfield  {journal}
  {\bibinfo  {journal} {Phys. Rev. Lett.}\ }\textbf {\bibinfo {volume} {127}},\
  \bibinfo {pages} {042303} (\bibinfo {year} {2021})}\BibitemShut {NoStop}%
\bibitem [{\citenamefont {Lu}, \citenamefont {Xia},\ and\ \citenamefont
  {Ruggieri}(2020)}]{Lu:2019diy}%
  \BibitemOpen
  \bibfield  {author} {\bibinfo {author} {\bibfnamefont {Z.-Y.}\ \bibnamefont
  {Lu}}, \bibinfo {author} {\bibfnamefont {C.-J.}\ \bibnamefont {Xia}},\ and\
  \bibinfo {author} {\bibfnamefont {M.}~\bibnamefont {Ruggieri}},\ }\href
  {https://doi.org/10.1140/epjc/s10052-020-7614-6} {\bibfield  {journal}
  {\bibinfo  {journal} {Eur. Phys. J. C}\ }\textbf {\bibinfo {volume} {80}},\
  \bibinfo {pages} {46} (\bibinfo {year} {2020})},\ \Eprint
  {https://arxiv.org/abs/1907.11497} {arXiv:1907.11497 [hep-ph]} \BibitemShut
  {NoStop}%
\bibitem [{\citenamefont {Karsch}\ \emph {et~al.}(2016)\citenamefont {Karsch}
  \emph {et~al.}}]{Karsch:2015nqx}%
  \BibitemOpen
  \bibfield  {author} {\bibinfo {author} {\bibfnamefont {F.}~\bibnamefont
  {Karsch}} \emph {et~al.},\ }\href
  {https://doi.org/10.1016/j.nuclphysa.2016.01.008} {\bibfield  {journal}
  {\bibinfo  {journal} {Nucl. Phys. A}\ }\textbf {\bibinfo {volume} {956}},\
  \bibinfo {pages} {352} (\bibinfo {year} {2016})}\BibitemShut {NoStop}%
\bibitem [{\citenamefont {Gavai}, \citenamefont {Gupta},\ and\ \citenamefont
  {Mukherjee}(2005)}]{Gavai:2004se}%
  \BibitemOpen
  \bibfield  {author} {\bibinfo {author} {\bibfnamefont {R.~V.}\ \bibnamefont
  {Gavai}}, \bibinfo {author} {\bibfnamefont {S.}~\bibnamefont {Gupta}},\ and\
  \bibinfo {author} {\bibfnamefont {S.}~\bibnamefont {Mukherjee}},\ }\href
  {https://doi.org/10.1103/PhysRevD.71.074013} {\bibfield  {journal} {\bibinfo
  {journal} {Phys. Rev. D}\ }\textbf {\bibinfo {volume} {71}},\ \bibinfo
  {pages} {074013} (\bibinfo {year} {2005})}\BibitemShut {NoStop}%
\bibitem [{\citenamefont {Borsanyi}\ \emph {et~al.}(2023)\citenamefont
  {Borsanyi}, \citenamefont {Fodor}, \citenamefont {Giordano}, \citenamefont
  {Guenther}, \citenamefont {Katz}, \citenamefont {Pasztor},\ and\
  \citenamefont {Wong}}]{Borsanyi:2022soo}%
  \BibitemOpen
  \bibfield  {author} {\bibinfo {author} {\bibfnamefont {S.}~\bibnamefont
  {Borsanyi}}, \bibinfo {author} {\bibfnamefont {Z.}~\bibnamefont {Fodor}},
  \bibinfo {author} {\bibfnamefont {M.}~\bibnamefont {Giordano}}, \bibinfo
  {author} {\bibfnamefont {J.~N.}\ \bibnamefont {Guenther}}, \bibinfo {author}
  {\bibfnamefont {S.~D.}\ \bibnamefont {Katz}}, \bibinfo {author}
  {\bibfnamefont {A.}~\bibnamefont {Pasztor}},\ and\ \bibinfo {author}
  {\bibfnamefont {C.~H.}\ \bibnamefont {Wong}},\ }\href
  {https://doi.org/10.1103/PhysRevD.107.L091503} {\bibfield  {journal}
  {\bibinfo  {journal} {Phys. Rev. D}\ }\textbf {\bibinfo {volume} {107}},\
  \bibinfo {pages} {L091503} (\bibinfo {year} {2023})},\ \Eprint
  {https://arxiv.org/abs/2208.05398} {arXiv:2208.05398 [hep-lat]} \BibitemShut
  {NoStop}%
\bibitem [{\citenamefont {Bazavov}\ \emph {et~al.}(2019)\citenamefont {Bazavov}
  \emph {et~al.}}]{HotQCD:2018pds}%
  \BibitemOpen
  \bibfield  {author} {\bibinfo {author} {\bibfnamefont {A.}~\bibnamefont
  {Bazavov}} \emph {et~al.} (\bibinfo {collaboration} {HotQCD}),\ }\href
  {https://doi.org/10.1016/j.physletb.2019.05.013} {\bibfield  {journal}
  {\bibinfo  {journal} {Phys. Lett. B}\ }\textbf {\bibinfo {volume} {795}},\
  \bibinfo {pages} {15} (\bibinfo {year} {2019})},\ \Eprint
  {https://arxiv.org/abs/1812.08235} {arXiv:1812.08235 [hep-lat]} \BibitemShut
  {NoStop}%
\bibitem [{\citenamefont {Borsanyi}\ \emph {et~al.}(2020)\citenamefont
  {Borsanyi}, \citenamefont {Fodor}, \citenamefont {Guenther}, \citenamefont
  {Kara}, \citenamefont {Katz}, \citenamefont {Parotto}, \citenamefont
  {Pasztor}, \citenamefont {Ratti},\ and\ \citenamefont
  {Szabo}}]{Borsanyi:2020fev}%
  \BibitemOpen
  \bibfield  {author} {\bibinfo {author} {\bibfnamefont {S.}~\bibnamefont
  {Borsanyi}}, \bibinfo {author} {\bibfnamefont {Z.}~\bibnamefont {Fodor}},
  \bibinfo {author} {\bibfnamefont {J.~N.}\ \bibnamefont {Guenther}}, \bibinfo
  {author} {\bibfnamefont {R.}~\bibnamefont {Kara}}, \bibinfo {author}
  {\bibfnamefont {S.~D.}\ \bibnamefont {Katz}}, \bibinfo {author}
  {\bibfnamefont {P.}~\bibnamefont {Parotto}}, \bibinfo {author} {\bibfnamefont
  {A.}~\bibnamefont {Pasztor}}, \bibinfo {author} {\bibfnamefont
  {C.}~\bibnamefont {Ratti}},\ and\ \bibinfo {author} {\bibfnamefont {K.~K.}\
  \bibnamefont {Szabo}},\ }\href
  {https://doi.org/10.1103/PhysRevLett.125.052001} {\bibfield  {journal}
  {\bibinfo  {journal} {Phys. Rev. Lett.}\ }\textbf {\bibinfo {volume} {125}},\
  \bibinfo {pages} {052001} (\bibinfo {year} {2020})}\BibitemShut {NoStop}%
\bibitem [{\citenamefont {Romatschke}\ and\ \citenamefont
  {Romatschke}(2007)}]{Romatschke:2007mq}%
  \BibitemOpen
  \bibfield  {author} {\bibinfo {author} {\bibfnamefont {P.}~\bibnamefont
  {Romatschke}}\ and\ \bibinfo {author} {\bibfnamefont {U.}~\bibnamefont
  {Romatschke}},\ }\href {https://doi.org/10.1103/PhysRevLett.99.172301}
  {\bibfield  {journal} {\bibinfo  {journal} {Phys. Rev. Lett.}\ }\textbf
  {\bibinfo {volume} {99}},\ \bibinfo {pages} {172301} (\bibinfo {year}
  {2007})}\BibitemShut {NoStop}%
\bibitem [{\citenamefont {Alba}\ \emph {et~al.}(2017)\citenamefont {Alba} \emph
  {et~al.}}]{Alba:2017mqu}%
  \BibitemOpen
  \bibfield  {author} {\bibinfo {author} {\bibfnamefont {P.}~\bibnamefont
  {Alba}} \emph {et~al.},\ }\href {https://doi.org/10.1103/PhysRevD.96.034517}
  {\bibfield  {journal} {\bibinfo  {journal} {Phys. Rev. D}\ }\textbf {\bibinfo
  {volume} {96}},\ \bibinfo {pages} {034517} (\bibinfo {year}
  {2017})}\BibitemShut {NoStop}%
\bibitem [{\citenamefont {Ma}\ \emph {et~al.}(2026)\citenamefont {Ma},
  \citenamefont {Guo}, \citenamefont {Chen}, \citenamefont {Liu}, \citenamefont
  {Zhu},\ and\ \citenamefont {Lu}}]{Ma:2026xxx}%
  \BibitemOpen
  \bibfield  {author} {\bibinfo {author} {\bibfnamefont {Z.-J.}\ \bibnamefont
  {Ma}}, \bibinfo {author} {\bibfnamefont {C.-G.}\ \bibnamefont {Guo}},
  \bibinfo {author} {\bibfnamefont {H.-M.}\ \bibnamefont {Chen}}, \bibinfo
  {author} {\bibfnamefont {L.-M.}\ \bibnamefont {Liu}}, \bibinfo {author}
  {\bibfnamefont {A.-N.}\ \bibnamefont {Zhu}},\ and\ \bibinfo {author}
  {\bibfnamefont {Z.-Y.}\ \bibnamefont {Lu}},\ }\href@noop {} {\bibfield
  {journal} {\bibinfo  {journal} {to appear}\ } (\bibinfo {year}
  {2026})}\BibitemShut {NoStop}%
\bibitem [{\citenamefont {Chen}, \citenamefont {Xia},\ and\ \citenamefont
  {Peng}(2022)}]{Chen:2021edy}%
  \BibitemOpen
  \bibfield  {author} {\bibinfo {author} {\bibfnamefont {H.-M.}\ \bibnamefont
  {Chen}}, \bibinfo {author} {\bibfnamefont {C.-J.}\ \bibnamefont {Xia}},\ and\
  \bibinfo {author} {\bibfnamefont {G.-X.}\ \bibnamefont {Peng}},\ }\href
  {https://doi.org/10.1088/1674-1137/ac4b5b} {\bibfield  {journal} {\bibinfo
  {journal} {Chin. Phys. C}\ }\textbf {\bibinfo {volume} {46}},\ \bibinfo
  {pages} {055102} (\bibinfo {year} {2022})},\ \Eprint
  {https://arxiv.org/abs/2110.09187} {arXiv:2110.09187 [nucl-th]} \BibitemShut
  {NoStop}%
\bibitem [{\citenamefont {Chen}, \citenamefont {Xia},\ and\ \citenamefont
  {Peng}(2024)}]{Chen:2023rza}%
  \BibitemOpen
  \bibfield  {author} {\bibinfo {author} {\bibfnamefont {H.-M.}\ \bibnamefont
  {Chen}}, \bibinfo {author} {\bibfnamefont {C.-J.}\ \bibnamefont {Xia}},\ and\
  \bibinfo {author} {\bibfnamefont {G.-X.}\ \bibnamefont {Peng}},\ }\href
  {https://doi.org/10.1103/PhysRevD.109.054031} {\bibfield  {journal} {\bibinfo
   {journal} {Phys. Rev. D}\ }\textbf {\bibinfo {volume} {109}},\ \bibinfo
  {pages} {054031} (\bibinfo {year} {2024})},\ \Eprint
  {https://arxiv.org/abs/2309.13583} {arXiv:2309.13583 [nucl-th]} \BibitemShut
  {NoStop}%
\bibitem [{\citenamefont {Wen}\ \emph {et~al.}(2005)\citenamefont {Wen},
  \citenamefont {Zhong}, \citenamefont {Peng}, \citenamefont {Shen},\ and\
  \citenamefont {Ning}}]{Wen:2005uf}%
  \BibitemOpen
  \bibfield  {author} {\bibinfo {author} {\bibfnamefont {X.~J.}\ \bibnamefont
  {Wen}}, \bibinfo {author} {\bibfnamefont {X.~H.}\ \bibnamefont {Zhong}},
  \bibinfo {author} {\bibfnamefont {G.~X.}\ \bibnamefont {Peng}}, \bibinfo
  {author} {\bibfnamefont {P.~N.}\ \bibnamefont {Shen}},\ and\ \bibinfo
  {author} {\bibfnamefont {P.~Z.}\ \bibnamefont {Ning}},\ }\href
  {https://doi.org/10.1103/PhysRevC.72.015204} {\bibfield  {journal} {\bibinfo
  {journal} {Phys. Rev. C}\ }\textbf {\bibinfo {volume} {72}},\ \bibinfo
  {pages} {015204} (\bibinfo {year} {2005})},\ \Eprint
  {https://arxiv.org/abs/hep-ph/0506050} {arXiv:hep-ph/0506050} \BibitemShut
  {NoStop}%
\bibitem [{\citenamefont {Xia}\ \emph {et~al.}(2014)\citenamefont {Xia},
  \citenamefont {Peng}, \citenamefont {Chen}, \citenamefont {Lu},\ and\
  \citenamefont {Xu}}]{Xia:2014zaa}%
  \BibitemOpen
  \bibfield  {author} {\bibinfo {author} {\bibfnamefont {C.~J.}\ \bibnamefont
  {Xia}}, \bibinfo {author} {\bibfnamefont {G.~X.}\ \bibnamefont {Peng}},
  \bibinfo {author} {\bibfnamefont {S.~W.}\ \bibnamefont {Chen}}, \bibinfo
  {author} {\bibfnamefont {Z.~Y.}\ \bibnamefont {Lu}},\ and\ \bibinfo {author}
  {\bibfnamefont {J.~F.}\ \bibnamefont {Xu}},\ }\href
  {https://doi.org/10.1103/PhysRevD.89.105027} {\bibfield  {journal} {\bibinfo
  {journal} {Phys. Rev. D}\ }\textbf {\bibinfo {volume} {89}},\ \bibinfo
  {pages} {105027} (\bibinfo {year} {2014})},\ \Eprint
  {https://arxiv.org/abs/1405.3037} {arXiv:1405.3037 [hep-ph]} \BibitemShut
  {NoStop}%
\bibitem [{\citenamefont {Lu}\ \emph {et~al.}(2016{\natexlab{a}})\citenamefont
  {Lu}, \citenamefont {Peng}, \citenamefont {Zhang}, \citenamefont {Ruggieri},\
  and\ \citenamefont {Greco}}]{Lu:2016jsv}%
  \BibitemOpen
  \bibfield  {author} {\bibinfo {author} {\bibfnamefont {Z.-Y.}\ \bibnamefont
  {Lu}}, \bibinfo {author} {\bibfnamefont {G.-X.}\ \bibnamefont {Peng}},
  \bibinfo {author} {\bibfnamefont {S.-P.}\ \bibnamefont {Zhang}}, \bibinfo
  {author} {\bibfnamefont {M.}~\bibnamefont {Ruggieri}},\ and\ \bibinfo
  {author} {\bibfnamefont {V.}~\bibnamefont {Greco}},\ }\href
  {https://doi.org/10.1007/s41365-016-0148-9} {\bibfield  {journal} {\bibinfo
  {journal} {Nucl. Sci. Tech.}\ }\textbf {\bibinfo {volume} {27}},\ \bibinfo
  {pages} {148} (\bibinfo {year} {2016}{\natexlab{a}})}\BibitemShut {NoStop}%
\bibitem [{\citenamefont {Issifu}(2024)}]{Issifu:2023qoo}%
  \BibitemOpen
  \bibfield  {author} {\bibinfo {author} {\bibfnamefont {A.}~\bibnamefont
  {Issifu}},\ }\href {https://doi.org/10.1140/epjc/s10052-024-12828-0}
  {\bibfield  {journal} {\bibinfo  {journal} {Eur. Phys. J. C}\ }\textbf
  {\bibinfo {volume} {84}},\ \bibinfo {pages} {463} (\bibinfo {year}
  {2024})}\BibitemShut {NoStop}%
\bibitem [{\citenamefont {Gorenstein}\ and\ \citenamefont
  {Yang}(1995)}]{Gorenstein:1995vm}%
  \BibitemOpen
  \bibfield  {author} {\bibinfo {author} {\bibfnamefont {M.~I.}\ \bibnamefont
  {Gorenstein}}\ and\ \bibinfo {author} {\bibfnamefont {S.-N.}\ \bibnamefont
  {Yang}},\ }\href {https://doi.org/10.1103/PhysRevD.52.5206} {\bibfield
  {journal} {\bibinfo  {journal} {Phys. Rev. D}\ }\textbf {\bibinfo {volume}
  {52}},\ \bibinfo {pages} {5206} (\bibinfo {year} {1995})}\BibitemShut
  {NoStop}%
\bibitem [{\citenamefont {Peshier}\ \emph {et~al.}(1996)\citenamefont
  {Peshier}, \citenamefont {Kampfer}, \citenamefont {Pavlenko},\ and\
  \citenamefont {Soff}}]{Peshier:1995ty}%
  \BibitemOpen
  \bibfield  {author} {\bibinfo {author} {\bibfnamefont {A.}~\bibnamefont
  {Peshier}}, \bibinfo {author} {\bibfnamefont {B.}~\bibnamefont {Kampfer}},
  \bibinfo {author} {\bibfnamefont {O.~P.}\ \bibnamefont {Pavlenko}},\ and\
  \bibinfo {author} {\bibfnamefont {G.}~\bibnamefont {Soff}},\ }\href
  {https://doi.org/10.1103/PhysRevD.54.2399} {\bibfield  {journal} {\bibinfo
  {journal} {Phys. Rev. D}\ }\textbf {\bibinfo {volume} {54}},\ \bibinfo
  {pages} {2399} (\bibinfo {year} {1996})}\BibitemShut {NoStop}%
\bibitem [{\citenamefont {Levai}\ and\ \citenamefont
  {Heinz}(1998)}]{Levai:1997yx}%
  \BibitemOpen
  \bibfield  {author} {\bibinfo {author} {\bibfnamefont {P.}~\bibnamefont
  {Levai}}\ and\ \bibinfo {author} {\bibfnamefont {U.~W.}\ \bibnamefont
  {Heinz}},\ }\href {https://doi.org/10.1103/PhysRevC.57.1879} {\bibfield
  {journal} {\bibinfo  {journal} {Phys. Rev. C}\ }\textbf {\bibinfo {volume}
  {57}},\ \bibinfo {pages} {1879} (\bibinfo {year} {1998})}\BibitemShut
  {NoStop}%
\bibitem [{\citenamefont {Bannur}(2007)}]{Bannur:2005wm}%
  \BibitemOpen
  \bibfield  {author} {\bibinfo {author} {\bibfnamefont {V.~M.}\ \bibnamefont
  {Bannur}},\ }\href {https://doi.org/10.1140/epjc/s10052-007-0233-7}
  {\bibfield  {journal} {\bibinfo  {journal} {Eur. Phys. J. C}\ }\textbf
  {\bibinfo {volume} {50}},\ \bibinfo {pages} {629} (\bibinfo {year} {2007})},\
  \Eprint {https://arxiv.org/abs/hep-ph/0508069} {arXiv:hep-ph/0508069}
  \BibitemShut {NoStop}%
\bibitem [{\citenamefont {Luo}\ \emph {et~al.}(2013)\citenamefont {Luo},
  \citenamefont {Cao}, \citenamefont {Yan}, \citenamefont {Sun},\ and\
  \citenamefont {Zong}}]{Luo:2013bcz}%
  \BibitemOpen
  \bibfield  {author} {\bibinfo {author} {\bibfnamefont {L.-J.}\ \bibnamefont
  {Luo}}, \bibinfo {author} {\bibfnamefont {J.}~\bibnamefont {Cao}}, \bibinfo
  {author} {\bibfnamefont {Y.}~\bibnamefont {Yan}}, \bibinfo {author}
  {\bibfnamefont {W.-M.}\ \bibnamefont {Sun}},\ and\ \bibinfo {author}
  {\bibfnamefont {H.-S.}\ \bibnamefont {Zong}},\ }\href
  {https://doi.org/10.1140/epjc/s10052-013-2626-0} {\bibfield  {journal}
  {\bibinfo  {journal} {Eur. Phys. J. C}\ }\textbf {\bibinfo {volume} {73}},\
  \bibinfo {pages} {2626} (\bibinfo {year} {2013})},\ \Eprint
  {https://arxiv.org/abs/1311.4017} {arXiv:1311.4017 [hep-ph]} \BibitemShut
  {NoStop}%
\bibitem [{\citenamefont {Sambataro}\ \emph {et~al.}(2024)\citenamefont
  {Sambataro}, \citenamefont {Greco}, \citenamefont {Parisi},\ and\
  \citenamefont {Plumari}}]{Sambataro:2024mkr}%
  \BibitemOpen
  \bibfield  {author} {\bibinfo {author} {\bibfnamefont {M.~L.}\ \bibnamefont
  {Sambataro}}, \bibinfo {author} {\bibfnamefont {V.}~\bibnamefont {Greco}},
  \bibinfo {author} {\bibfnamefont {G.}~\bibnamefont {Parisi}},\ and\ \bibinfo
  {author} {\bibfnamefont {S.}~\bibnamefont {Plumari}},\ }\href
  {https://doi.org/10.1140/epjc/s10052-024-13276-6} {\bibfield  {journal}
  {\bibinfo  {journal} {Eur. Phys. J. C}\ }\textbf {\bibinfo {volume} {84}},\
  \bibinfo {pages} {881} (\bibinfo {year} {2024})}\BibitemShut {NoStop}%
\bibitem [{\citenamefont {Ma}\ \emph {et~al.}(2023)\citenamefont {Ma},
  \citenamefont {Lu}, \citenamefont {Xu}, \citenamefont {Peng}, \citenamefont
  {Fu},\ and\ \citenamefont {Wang}}]{Ma:2023stj}%
  \BibitemOpen
  \bibfield  {author} {\bibinfo {author} {\bibfnamefont {Z.-J.}\ \bibnamefont
  {Ma}}, \bibinfo {author} {\bibfnamefont {Z.-Y.}\ \bibnamefont {Lu}}, \bibinfo
  {author} {\bibfnamefont {J.-F.}\ \bibnamefont {Xu}}, \bibinfo {author}
  {\bibfnamefont {G.-X.}\ \bibnamefont {Peng}}, \bibinfo {author}
  {\bibfnamefont {X.}~\bibnamefont {Fu}},\ and\ \bibinfo {author}
  {\bibfnamefont {J.}~\bibnamefont {Wang}},\ }\href
  {https://doi.org/10.1103/PhysRevD.108.054017} {\bibfield  {journal} {\bibinfo
   {journal} {Phys. Rev. D}\ }\textbf {\bibinfo {volume} {108}},\ \bibinfo
  {pages} {054017} (\bibinfo {year} {2023})},\ \Eprint
  {https://arxiv.org/abs/2308.05308} {arXiv:2308.05308 [hep-ph]} \BibitemShut
  {NoStop}%
\bibitem [{\citenamefont {Wang}\ \emph {et~al.}(2025)\citenamefont {Wang},
  \citenamefont {Lu}, \citenamefont {Ma}, \citenamefont {Yang}, \citenamefont
  {Xu},\ and\ \citenamefont {Fu}}]{Wang:2025lwv}%
  \BibitemOpen
  \bibfield  {author} {\bibinfo {author} {\bibfnamefont {S.-P.}\ \bibnamefont
  {Wang}}, \bibinfo {author} {\bibfnamefont {Z.-Y.}\ \bibnamefont {Lu}},
  \bibinfo {author} {\bibfnamefont {Z.-J.}\ \bibnamefont {Ma}}, \bibinfo
  {author} {\bibfnamefont {R.-Y.}\ \bibnamefont {Yang}}, \bibinfo {author}
  {\bibfnamefont {J.-F.}\ \bibnamefont {Xu}},\ and\ \bibinfo {author}
  {\bibfnamefont {X.}~\bibnamefont {Fu}},\ }\href
  {https://doi.org/10.1103/d7pv-1dqt} {\bibfield  {journal} {\bibinfo
  {journal} {Phys. Rev. D}\ }\textbf {\bibinfo {volume} {112}},\ \bibinfo
  {pages} {094036} (\bibinfo {year} {2025})}\BibitemShut {NoStop}%
\bibitem [{\citenamefont {Bannur}(2008)}]{Bannur:2007tk}%
  \BibitemOpen
  \bibfield  {author} {\bibinfo {author} {\bibfnamefont {V.~M.}\ \bibnamefont
  {Bannur}},\ }\href {https://doi.org/10.1103/PhysRevC.78.045206} {\bibfield
  {journal} {\bibinfo  {journal} {Phys. Rev. C}\ }\textbf {\bibinfo {volume}
  {78}},\ \bibinfo {pages} {045206} (\bibinfo {year} {2008})},\ \Eprint
  {https://arxiv.org/abs/0712.2886} {arXiv:0712.2886 [hep-ph]} \BibitemShut
  {NoStop}%
\bibitem [{\citenamefont {Peng}(2006)}]{Peng:2006tk}%
  \BibitemOpen
  \bibfield  {author} {\bibinfo {author} {\bibfnamefont {G.~X.}\ \bibnamefont
  {Peng}},\ }\href {https://doi.org/10.1016/j.physletb.2006.01.032} {\bibfield
  {journal} {\bibinfo  {journal} {Phys. Lett. B}\ }\textbf {\bibinfo {volume}
  {634}},\ \bibinfo {pages} {413} (\bibinfo {year} {2006})},\ \Eprint
  {https://arxiv.org/abs/hep-th/0601051} {arXiv:hep-th/0601051} \BibitemShut
  {NoStop}%
\bibitem [{\citenamefont {Lu}\ \emph {et~al.}(2016{\natexlab{b}})\citenamefont
  {Lu}, \citenamefont {Peng}, \citenamefont {Xu},\ and\ \citenamefont
  {Zhang}}]{Lu:2016fki}%
  \BibitemOpen
  \bibfield  {author} {\bibinfo {author} {\bibfnamefont {Z.-Y.}\ \bibnamefont
  {Lu}}, \bibinfo {author} {\bibfnamefont {G.-X.}\ \bibnamefont {Peng}},
  \bibinfo {author} {\bibfnamefont {J.-F.}\ \bibnamefont {Xu}},\ and\ \bibinfo
  {author} {\bibfnamefont {S.-P.}\ \bibnamefont {Zhang}},\ }\href
  {https://doi.org/10.1007/s11433-015-0524-2} {\bibfield  {journal} {\bibinfo
  {journal} {Sci. China Phys. Mech. Astron.}\ }\textbf {\bibinfo {volume}
  {59}},\ \bibinfo {pages} {662001} (\bibinfo {year}
  {2016}{\natexlab{b}})}\BibitemShut {NoStop}%
\bibitem [{\citenamefont {Buisseret}(2010)}]{Buisseret:2010mop}%
  \BibitemOpen
  \bibfield  {author} {\bibinfo {author} {\bibfnamefont {F.}~\bibnamefont
  {Buisseret}},\ }\href {https://doi.org/10.1140/epjc/s10052-010-1341-3}
  {\bibfield  {journal} {\bibinfo  {journal} {Eur. Phys. J. C}\ }\textbf
  {\bibinfo {volume} {68}},\ \bibinfo {pages} {473} (\bibinfo {year}
  {2010})}\BibitemShut {NoStop}%
\bibitem [{\citenamefont {Trotti}, \citenamefont {Jafarzade},\ and\
  \citenamefont {Giacosa}(2023)}]{Trotti:2022knd}%
  \BibitemOpen
  \bibfield  {author} {\bibinfo {author} {\bibfnamefont {E.}~\bibnamefont
  {Trotti}}, \bibinfo {author} {\bibfnamefont {S.}~\bibnamefont {Jafarzade}},\
  and\ \bibinfo {author} {\bibfnamefont {F.}~\bibnamefont {Giacosa}},\ }\href
  {https://doi.org/10.1140/epjc/s10052-023-11557-0} {\bibfield  {journal}
  {\bibinfo  {journal} {Eur. Phys. J. C}\ }\textbf {\bibinfo {volume} {83}},\
  \bibinfo {pages} {390} (\bibinfo {year} {2023})}\BibitemShut {NoStop}%
\bibitem [{\citenamefont {Meisinger}, \citenamefont {Miller},\ and\
  \citenamefont {Ogilvie}(2002)}]{Meisinger:2001cq}%
  \BibitemOpen
  \bibfield  {author} {\bibinfo {author} {\bibfnamefont {P.~N.}\ \bibnamefont
  {Meisinger}}, \bibinfo {author} {\bibfnamefont {T.~R.}\ \bibnamefont
  {Miller}},\ and\ \bibinfo {author} {\bibfnamefont {M.~C.}\ \bibnamefont
  {Ogilvie}},\ }\href {https://doi.org/10.1103/PhysRevD.65.034009} {\bibfield
  {journal} {\bibinfo  {journal} {Phys. Rev. D}\ }\textbf {\bibinfo {volume}
  {65}},\ \bibinfo {pages} {034009} (\bibinfo {year} {2002})},\ \Eprint
  {https://arxiv.org/abs/hep-ph/0108009} {arXiv:hep-ph/0108009} \BibitemShut
  {NoStop}%
\bibitem [{\citenamefont {Sasaki}, \citenamefont {Mishustin},\ and\
  \citenamefont {Redlich}(2014)}]{Sasaki:2013xfa}%
  \BibitemOpen
  \bibfield  {author} {\bibinfo {author} {\bibfnamefont {C.}~\bibnamefont
  {Sasaki}}, \bibinfo {author} {\bibfnamefont {I.}~\bibnamefont {Mishustin}},\
  and\ \bibinfo {author} {\bibfnamefont {K.}~\bibnamefont {Redlich}},\ }\href
  {https://doi.org/10.1103/PhysRevD.89.014031} {\bibfield  {journal} {\bibinfo
  {journal} {Phys. Rev. D}\ }\textbf {\bibinfo {volume} {89}},\ \bibinfo
  {pages} {014031} (\bibinfo {year} {2014})}\BibitemShut {NoStop}%
\bibitem [{\citenamefont {Lo}\ \emph {et~al.}(2013)\citenamefont {Lo},
  \citenamefont {Friman}, \citenamefont {Kaczmarek}, \citenamefont {Redlich},\
  and\ \citenamefont {Sasaki}}]{Lo:2013hla}%
  \BibitemOpen
  \bibfield  {author} {\bibinfo {author} {\bibfnamefont {P.~M.}\ \bibnamefont
  {Lo}}, \bibinfo {author} {\bibfnamefont {B.}~\bibnamefont {Friman}}, \bibinfo
  {author} {\bibfnamefont {O.}~\bibnamefont {Kaczmarek}}, \bibinfo {author}
  {\bibfnamefont {K.}~\bibnamefont {Redlich}},\ and\ \bibinfo {author}
  {\bibfnamefont {C.}~\bibnamefont {Sasaki}},\ }\href
  {https://doi.org/10.1103/PhysRevD.88.074502} {\bibfield  {journal} {\bibinfo
  {journal} {Phys. Rev. D}\ }\textbf {\bibinfo {volume} {88}},\ \bibinfo
  {pages} {074502} (\bibinfo {year} {2013})}\BibitemShut {NoStop}%
\bibitem [{\citenamefont {Singha}\ \emph {et~al.}(2022)\citenamefont {Singha},
  \citenamefont {Ray}, \citenamefont {Islam},\ and\ \citenamefont
  {Mustafa}}]{Singha:2021wpi}%
  \BibitemOpen
  \bibfield  {author} {\bibinfo {author} {\bibfnamefont {P.}~\bibnamefont
  {Singha}}, \bibinfo {author} {\bibfnamefont {R.}~\bibnamefont {Ray}},
  \bibinfo {author} {\bibfnamefont {C.~A.}\ \bibnamefont {Islam}},\ and\
  \bibinfo {author} {\bibfnamefont {M.~G.}\ \bibnamefont {Mustafa}},\ }\href
  {https://doi.org/10.1007/978-981-19-2354-8_77} {\bibfield  {journal}
  {\bibinfo  {journal} {Springer Proc. Phys.}\ }\textbf {\bibinfo {volume}
  {277}},\ \bibinfo {pages} {423} (\bibinfo {year} {2022})}\BibitemShut
  {NoStop}%
\bibitem [{\citenamefont {Motta}\ \emph {et~al.}(2020)\citenamefont {Motta},
  \citenamefont {Stiele}, \citenamefont {Alberico},\ and\ \citenamefont
  {Beraudo}}]{Motta:2020cbr}%
  \BibitemOpen
  \bibfield  {author} {\bibinfo {author} {\bibfnamefont {M.}~\bibnamefont
  {Motta}}, \bibinfo {author} {\bibfnamefont {R.}~\bibnamefont {Stiele}},
  \bibinfo {author} {\bibfnamefont {W.~M.}\ \bibnamefont {Alberico}},\ and\
  \bibinfo {author} {\bibfnamefont {A.}~\bibnamefont {Beraudo}},\ }\href
  {https://doi.org/10.1140/epjc/s10052-020-8218-x} {\bibfield  {journal}
  {\bibinfo  {journal} {Eur. Phys. J. C}\ }\textbf {\bibinfo {volume} {80}},\
  \bibinfo {pages} {770} (\bibinfo {year} {2020})}\BibitemShut {NoStop}%
\bibitem [{\citenamefont {Khaidukov}, \citenamefont {Lukashov},\ and\
  \citenamefont {Simonov}(2018)}]{Khaidukov:2018lor}%
  \BibitemOpen
  \bibfield  {author} {\bibinfo {author} {\bibfnamefont {Z.~V.}\ \bibnamefont
  {Khaidukov}}, \bibinfo {author} {\bibfnamefont {M.~S.}\ \bibnamefont
  {Lukashov}},\ and\ \bibinfo {author} {\bibfnamefont {{\relax Yu}.~A.}\
  \bibnamefont {Simonov}},\ }\href {https://doi.org/10.1103/PhysRevD.98.074031}
  {\bibfield  {journal} {\bibinfo  {journal} {Phys. Rev. D}\ }\textbf {\bibinfo
  {volume} {98}},\ \bibinfo {pages} {074031} (\bibinfo {year} {2018})},\
  \Eprint {https://arxiv.org/abs/1806.09407} {arXiv:1806.09407 [hep-ph]}
  \BibitemShut {NoStop}%
\bibitem [{\citenamefont {Bollweg}\ \emph {et~al.}(2023)\citenamefont
  {Bollweg}, \citenamefont {Clarke}, \citenamefont {Goswami}, \citenamefont
  {Kaczmarek}, \citenamefont {Karsch}, \citenamefont {Mukherjee}, \citenamefont
  {Petreczky}, \citenamefont {Schmidt},\ and\ \citenamefont
  {Sharma}}]{Bollweg:2022fqq}%
  \BibitemOpen
  \bibfield  {author} {\bibinfo {author} {\bibfnamefont {D.}~\bibnamefont
  {Bollweg}}, \bibinfo {author} {\bibfnamefont {D.~A.}\ \bibnamefont {Clarke}},
  \bibinfo {author} {\bibfnamefont {J.}~\bibnamefont {Goswami}}, \bibinfo
  {author} {\bibfnamefont {O.}~\bibnamefont {Kaczmarek}}, \bibinfo {author}
  {\bibfnamefont {F.}~\bibnamefont {Karsch}}, \bibinfo {author} {\bibfnamefont
  {S.}~\bibnamefont {Mukherjee}}, \bibinfo {author} {\bibfnamefont
  {P.}~\bibnamefont {Petreczky}}, \bibinfo {author} {\bibfnamefont
  {C.}~\bibnamefont {Schmidt}},\ and\ \bibinfo {author} {\bibfnamefont
  {S.}~\bibnamefont {Sharma}},\ }\href
  {https://doi.org/10.1103/PhysRevD.108.014510} {\bibfield  {journal} {\bibinfo
   {journal} {Phys. Rev. D}\ }\textbf {\bibinfo {volume} {108}},\ \bibinfo
  {pages} {014510} (\bibinfo {year} {2023})}\BibitemShut {NoStop}%
\bibitem [{\citenamefont {Rath}\ and\ \citenamefont
  {Dash}(2022)}]{Rath:2022oum}%
  \BibitemOpen
  \bibfield  {author} {\bibinfo {author} {\bibfnamefont {S.}~\bibnamefont
  {Rath}}\ and\ \bibinfo {author} {\bibfnamefont {S.}~\bibnamefont {Dash}},\
  }\href {https://doi.org/10.1140/epjc/s10052-022-10757-4} {\bibfield
  {journal} {\bibinfo  {journal} {Eur. Phys. J. C}\ }\textbf {\bibinfo {volume}
  {82}},\ \bibinfo {pages} {797} (\bibinfo {year} {2022})}\BibitemShut
  {NoStop}%
\bibitem [{\citenamefont {Frasc{\`a}}, \citenamefont {Beraudo},\ and\
  \citenamefont {Del~Zanna}(2025)}]{Frasca:2025ffn}%
  \BibitemOpen
  \bibfield  {author} {\bibinfo {author} {\bibfnamefont {F.}~\bibnamefont
  {Frasc{\`a}}}, \bibinfo {author} {\bibfnamefont {A.}~\bibnamefont
  {Beraudo}},\ and\ \bibinfo {author} {\bibfnamefont {L.}~\bibnamefont
  {Del~Zanna}},\ }\href {https://doi.org/10.1140/epjc/s10052-025-14955-8}
  {\bibfield  {journal} {\bibinfo  {journal} {Eur. Phys. J. C}\ }\textbf
  {\bibinfo {volume} {85}},\ \bibinfo {pages} {1202} (\bibinfo {year}
  {2025})}\BibitemShut {NoStop}%
\bibitem [{\citenamefont {Plumari}\ \emph {et~al.}(2012)\citenamefont
  {Plumari}, \citenamefont {Puglisi}, \citenamefont {Scardina},\ and\
  \citenamefont {Greco}}]{Plumari:2012ep}%
  \BibitemOpen
  \bibfield  {author} {\bibinfo {author} {\bibfnamefont {S.}~\bibnamefont
  {Plumari}}, \bibinfo {author} {\bibfnamefont {A.}~\bibnamefont {Puglisi}},
  \bibinfo {author} {\bibfnamefont {F.}~\bibnamefont {Scardina}},\ and\
  \bibinfo {author} {\bibfnamefont {V.}~\bibnamefont {Greco}},\ }\href
  {https://doi.org/10.1103/PhysRevC.86.054902} {\bibfield  {journal} {\bibinfo
  {journal} {Phys. Rev. C}\ }\textbf {\bibinfo {volume} {86}},\ \bibinfo
  {pages} {054902} (\bibinfo {year} {2012})},\ \Eprint
  {https://arxiv.org/abs/1208.0481} {arXiv:1208.0481 [nucl-th]} \BibitemShut
  {NoStop}%
\end{thebibliography}%

\end{document}